\documentclass[%
 reprint,
 amsmath,amssymb,
 aps,
]{revtex4-2}

\usepackage{graphicx}
\usepackage{dcolumn}
\usepackage{bm}
\usepackage{gensymb}

\begin{document}
\preprint{APS/123-QED}
\title{Local power coupling as a predictor of high-gradient breakdown performance}
\author{Jan Paszkiewicz}
\email{jan.paszkiewicz@cern.ch}
\author{Alexej Grudiev}
\author{Walter Wuensch}
\affiliation{CERN, CH-1211, Geneva 23, Switzerland}
\date{\today}

\begin{abstract}
A novel quantity for predicting the high-gradient performance of radio frequency accelerating structures is presented. The quantity is motivated, derived and compared with earlier high-gradient limits and experiments. This new method models a nascent RF breakdown as a current-carrying antenna and calculates the coupling of the antenna to an RF power source. With the help of an electron emission model to describe a nascent breakdown, the antenna model describes how a breakdown modifies the local surface electric field before it fully develops in any given structure geometry. For the structure geometries that this method was applied to, it was found that the calculated breakdown-loaded electric field was well-correlated with observed spatial breakdown distributions, and gave consistent values for the maximum breakdown-limited accelerating gradient between different geometries.
\end{abstract}

\maketitle

\section{\label{sec:intro}Introduction}
Predicting the high-gradient performance limit imposed by vacuum breakdown is an important part of the design process of many high-voltage vacuum components such as radio frequency particle accelerator structures. The breakdown limit is defined as the highest RF power, with corresponding field levels, at which a structure can be operated stably. Stability in high-gradient applications is often defined by a maximum breakdown rate, where the rate is given by number of breakdowns per pulse. To minimise the need for multiple iterations of designing, building, and testing, a reliable physical model for RF breakdown levels is an essential design tool for high-gradient linear accelerator applications.

Criteria involving the surface electric field $E$, such as Kilpatrick's criterion \cite{KilpatrickOriginal}, magnetic field $H$ \cite{PerformanceLimitingEffects}, or both, such as the ratio of power to iris circumference $P/C$ \cite{Grudiev2009}, or the modified Poynting vector $S_c$ \cite{Grudiev2009}, quantifying breakdown performance already exist, but their range of applicability is limited and are consistent with only certain measurements. Evidence is presented in Section \ref{sec:crab_results} for inconsistency between the predictions of the modified Poynting vector and experimental results. The new model proposed in this paper aims to be more general with a more complete fundamental physical basis than previous work and apply to a broader range of experimental results, whilst still being relatively straightforward in concept and simple to compute. In this paper, existing models will be reviewed and inconsistencies discussed in Sec. \ref{sec:review}, and the new model detailed in \ref{sec:proposal}. The results of the model will be explored for generic RF structure geometries in Sec.\ref{sec:c3}, and for some experimentally-tested exampled in Sec. \ref{sec:comparison_with_experiments}. Additional aspects of the model and its properties will be discussed in Sec.\ref{sec:extensions}, followed by its implications on optimal structure design in Sec. \ref{sec:implications}.

\section{\label{sec:review}Review of Existing Models}
It has been shown that a breakdown criterion that is purely a function of the surface electric field magnitude (such as Boyd's formulation of Kilpatrick's criterion often used in the field of accelerator physics \cite{KilpatrickBoyd}) does not fully describe the breakdown behaviour of accelerating structures, as experiments have shown that structures made of the same material and operating at the same RF frequency, but with different geometries, can reach very different peak surface electric field values. In addition to electric-field-based limits, alternatives based on global and local power flow as well as peak magnetic field and associated pulsed surface heating have been proposed \cite{PerformanceLimitingEffects}.

$S_c$ and $P/C$ depend on the RF power flowing through the structure rather than the electric field. They have been shown to produce predictions of maximum gradient that are more consistent with experimental results than the peak surface electric field alone. $P/C$, however, does not take into account the exact shape of the structure and thus cannot provide detailed insight into how to optimise the geometry for breakdown performance. It is also not applicable to standing-wave structures, in which the real power flow is low.

\begin{figure}[htbp]
\includegraphics[width=0.35\textwidth]{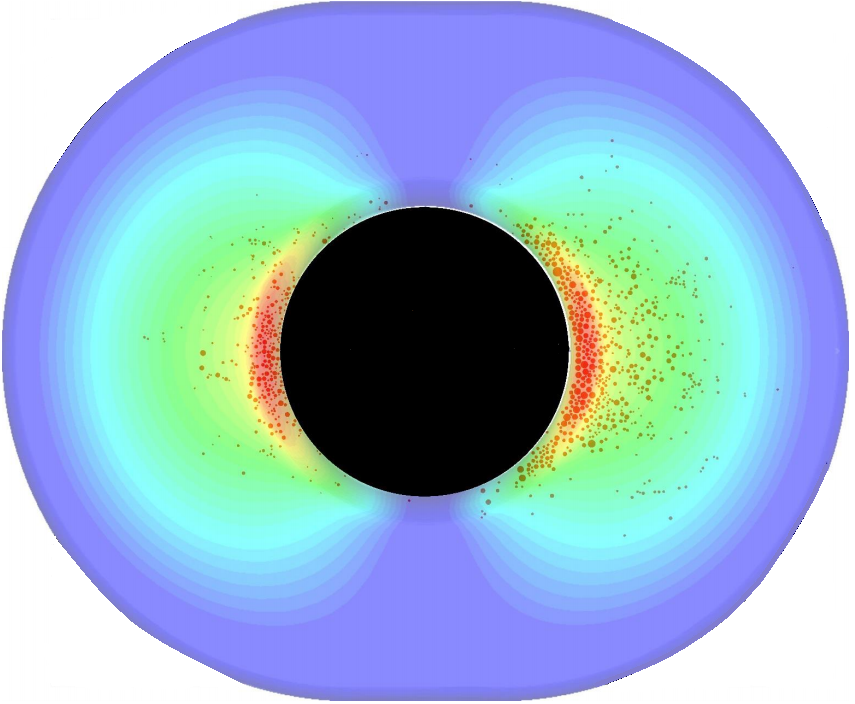}
\includegraphics[width=0.1\textwidth]{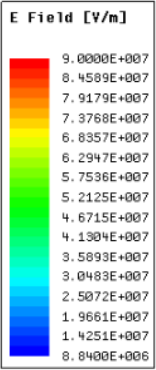}
\caption{\label{fig:crab_post_mortem_e}Breakdown locations, represented by black dots, vs. position transverse to the beam in the second cell of the CLIC Crab Cavity, obtained from \emph{post-mortem} analysis \cite{CrabPostMortem}. Colours represent the surface electric field vs. position transverse to the beam. The magnetic field has a similar distribution, but rotated by 90\degree compared to the electric field.}
\end{figure}

\begin{figure}[htbp]
\includegraphics[width=0.35\textwidth]{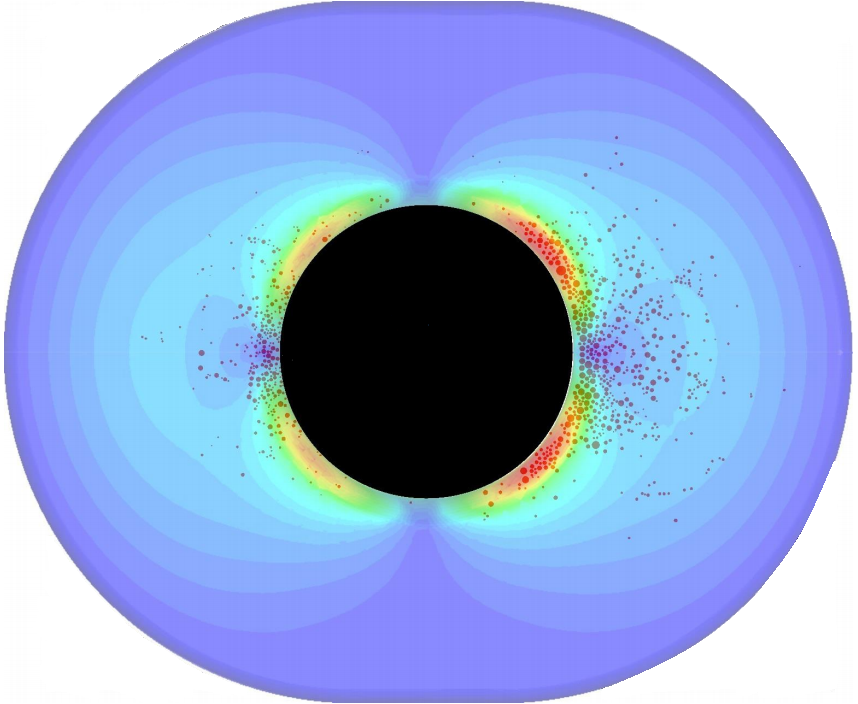}
\hspace{0.02\textwidth}
\includegraphics[width=0.1\textwidth]{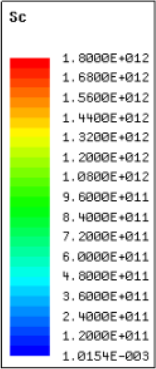}
\caption{\label{fig:crab_post_mortem_sc}Breakdown locations, represented by black dots, vs. position transverse to the beam in the second cell of the CLIC Crab Cavity, obtained from \emph{post-mortem} analysis \cite{CrabPostMortem}. Colours represent $S_c$ value vs. position transverse to the beam.}
\end{figure}

Tests in the XBox test stands at CERN \cite{Woolley2015HighCavity} have shown cases in which the strongest clustering of breakdown craters was found to be in distinctly different locations from those predicted by the spatial distribution of $S_c$, and also magnetic field. One such case was the Compact Linear Collider (CLIC) Crab Cavity prototype \cite{CrabCavity_Report}, a backward travelling-wave constant-impedance structure with a group velocity of $0.029c$. Being a deflecting cavity, it was operated in the $TM_{110}$ mode as opposed to the $TM_{010}$ mode as would typically be the case with accelerating structures. This mode produces distinctly different electric field, magnetic field, and $S_c$ patterns. As can be seen in Figs.~\ref{fig:crab_post_mortem_e} and \ref{fig:crab_post_mortem_sc}, the spatial distribution of the breakdown craters observed in a \emph{post-mortem} study matched the spatial distribution of the surface electric field magnitude much more closely than that of $S_c$ and magnetic field \cite{CrabCavity_Report}. A large number of breakdowns even occurred at a location at which the $S_c$ value was close to zero. This signifies that neither $S_c$ nor magnetic field fully encapsulates the relationship between structure geometry and breakdown performance. Fig.~\ref{fig:crab_post_mortem_e} also shows that the breakdowns did not occur in locations of high magnetic field, which is inconsistent with models based on magnetic field and pulsed heating. The maximum input power and corresponding surface fields were, however, roughly in line with those predicted by $S_c$. The inconsistency in predicted breakdown locations motivates revisiting the question of high-gradient limits. In addition, there has been an improved understanding of the mechanisms of vacuum arcs which will be used to get a better limit.
Another shortcoming of Kilpatrick's criterion, $S_c$ and $H$ as predictors of breakdown performance is that they cannot be applied to direct-current (DC) setups, in which there is no steady-state power flow or surface currents. This implies that breakdowns should never occur in such systems under steady-state conditions, which is contrary to observations.

Nevertheless, $S_c$ has seen success in predicting the field levels in accelerating structures operating in the $TM_{010}$ mode and has become a widely used quantity in RF design as evidenced by the numerous citations of \cite{Grudiev2009}, suggesting that available power does play an important role in determining the breakdown limit.

\section{\label{sec:proposal}Loaded Electric Field Model}
\subsection{\label{sec:outline}Outline and Reasoning}
The proposed new model is intended to encompass the idea of power flow determining breakdown rate whilst also resolving the discrepancy between the location of the maximum $S_c$ value and that of breakdown craters discussed in Sec. \ref{sec:review}.

It is believed that an RF breakdown begins with a field-emission site \cite{KovermannThesis}, which is likely a nanometer-scale feature on the surface of an accelerating cavity subject to high electric field. These sites are believed to have a field-enhancing geometric protrusion, locally lower workfunction, or other characteristic that causes them to emit a greater field-emission current density than the surrounding area. At any given time, there are many field-emission sites inside a structure that may suddenly begin a runaway process in which the emitted current grows by many orders of magnitude due to the increase in temperature of the emitter and formation of a plasma spot and develop into a breakdown.

Under the new model, it is proposed that, depending on the amount of RF power available, a field-emission site that has begun this thermal runaway process may either fully develop into an observable breakdown or extinguish itself partway. It is assumed that since the event in the latter case is very short in duration and results in a relatively small perturbation in the RF field compared to a normal breakdown, it is not normally noticed in normal operation, since the field do not collapse, reflected power is not produced, and stored energy is not discharged. The method is an attempt to quantify this threshold, by determining the surface electric field at the breakdown site when loaded by the rapidly increasing electron current associated with the developing breakdown. The quasi-equilibrium loaded and unloaded electric field amplitudes at the location of the breakdown site are denoted $E^*$ and $E_0$ respectively. Due to the loading effect of the electrons, $E^* \leq E_0$. The reduction in electric field caused by the loading depends on the electromagnetic power available from the power source, with less power resulting in a greater reduction in field. This means that breakdowns are most likely to occur in locations with a large surface electric field and high power flow, but are unlikely to occur in locations with a low surface electric field even if the power flow is very large.

$E^*$ therefore determines the breakdown performance of a given geometry, and should not exceed a given threshold value determined by the maximum allowed breakdown rate anywhere in the geometry, analogously to $S_c$. The calculation of $E^*$ for a particular breakdown site involves an equivalent circuit model in which the breakdown site is represented by a load impedance and the RF structure and power source are represented by a voltage source with a series impedance $\hat{Z}_{bd}$. For convenience, the calculation is performed in terms of electric field rather than voltage. The coupling of RF power to the breakdown is modelled by a short monopole antenna located on the surface of the structure at the location of the breakdown. The relationship between the surface electric field $\hat{E}$ at the breakdown site and the emitted current $\hat{I}$, is thus given by:

\begin{equation} \label{eqn:e_loading_function_complex}
    \hat{E} = \hat{E_0} - \hat{I}\hat{Z}_{bd}/l_{ant},
\end{equation}

where $l_{ant}$ is the antenna length. Here, $\hat{E}$, $\hat{E_0}$, $\hat{I}$, and $\hat{Z}_{bd}$ are complex phasor quantities. The emitted current is likely a nonlinear function of the surface electric field. Therefore, if the applied electric field varies sinusoidally in time, as is typical for resonant structures, the emitted current will have a periodic waveform in phase with the electric field. In this paper, only the fundamental frequency component of this waveform will be considered, since this component is likely to have the greatest effect on the peak loaded electric field. Calculating the loaded electric field is important as the basic premise of the model is that progression to full breakdown can only occur if there is a sufficiently high electric field remaining with the loading present. Since we are concerned with the reduction in peak electric field, only the real part of the electric field perturbation is relevant. Given that $\hat{E}$ and $\hat{I}$ have already been assumed to be in phase, only the real part of $\hat{Z}_{bd}$ is needed. Eq.\eqref{eqn:e_loading_function_complex} can thus be reduced to the scalar equation:

\begin{equation} \label{eqn:e_loading_function}
    E = E_0 - IR_{bd}/l_{ant},
\end{equation}

which will be referred to as the load line, with $R_{bd}$ being the real part of $\hat{Z}_{bd}$, and $E$, $E_0$ and $I$ being scalar values indicating the magnitude of $\hat{E}$, $\hat{E_0}$, and $\hat{I}$ respectively. $R_{bd}$ can be formally defined as:

\begin{equation} \label{eqn:bd_impedance_approx}
    R_{bd} = \Re(\hat{Z}_{bd}) \approx \Re\left(\frac{\Delta E l_{ant}}{\Delta I}\right),
\end{equation}

where $\Delta E$ is the perturbation in electric field resulting from a current $\Delta I$ in the antenna. Because of the linearity of Maxwell's equations in vacuum \cite{MicrowaveEngineering}, and if the geometry of the antenna is not affected by the Lorentz force, $\hat{Z}_{bd}$ is independent of the field levels. Given an emission function $I = f(E)$ that describes the load impedance presented by the breakdown, $E^*$ can be determined by solving:

\begin{equation} \label{eqn:e_loading_solution}
    E^* = E_0 - f(E^*)R_{bd}/l_{ant}
\end{equation}

\begin{figure}[htbp]
\includegraphics[width=0.5\textwidth]{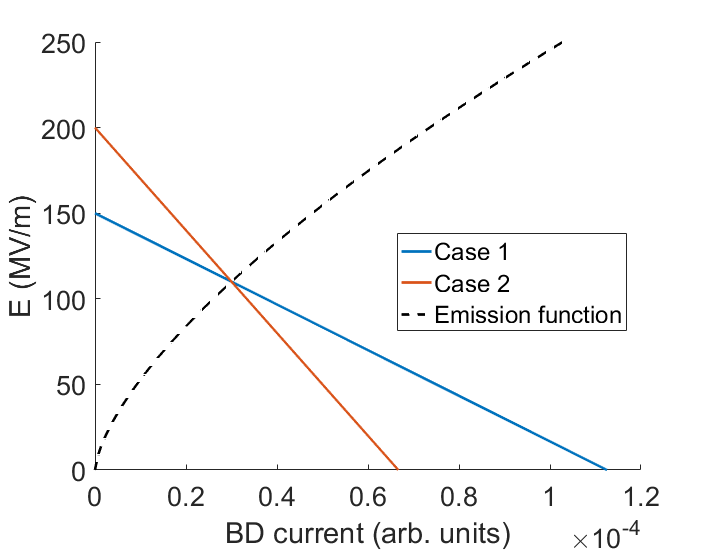}
\hspace{0.02\textwidth}
\caption{\label{fig:example_impedance}Calculation of $E^*$ on a plot of surface electric field in MV/m vs. current emitted by a breakdown. Black: emission function as given by Eq. \eqref{eqn:child_langmuir_emission}. Blue: a breakdown site with an unloaded field of 150 MV/m and a low breakdown resistance. Red: a breakdown site with an unloaded field of 200 MV/m and a larger breakdown resistance. }
\end{figure}

An example of a solution of Eq.\eqref{eqn:e_loading_solution} is shown in Fig.~\ref{fig:example_impedance}. Two potential breakdown sites are shown, whose load lines cross the emission function at the same point, resulting in the $E^*$ being the same for both, implying equal breakdown probability. However, due to the larger $R_{bd}$ associated with Case 2, the unloaded surface field $E_0$ is larger than that of Case 1. This example illustrates how, under the loaded-field model, different geometries can achieve different peak field levels for a given breakdown rate.

\begin{figure}[htbp]
\centering
\includegraphics[width=0.155\textwidth]{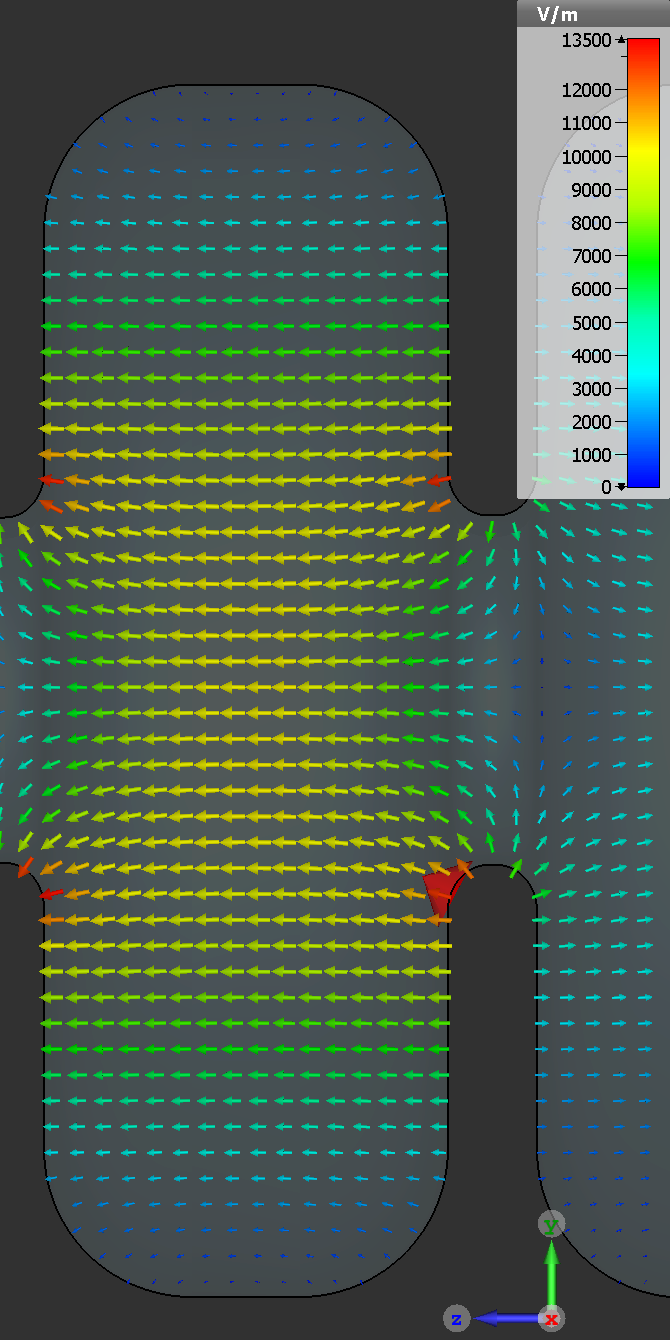}
\includegraphics[width=0.155\textwidth]{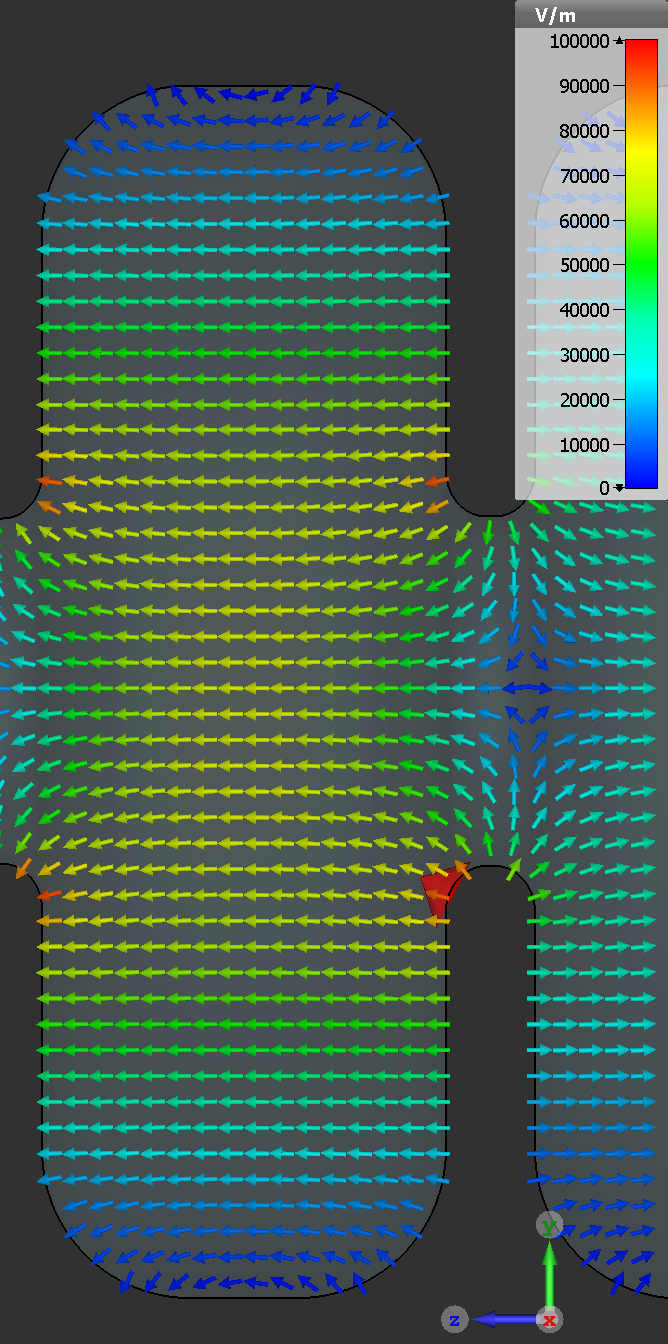}
\includegraphics[width=0.155\textwidth]{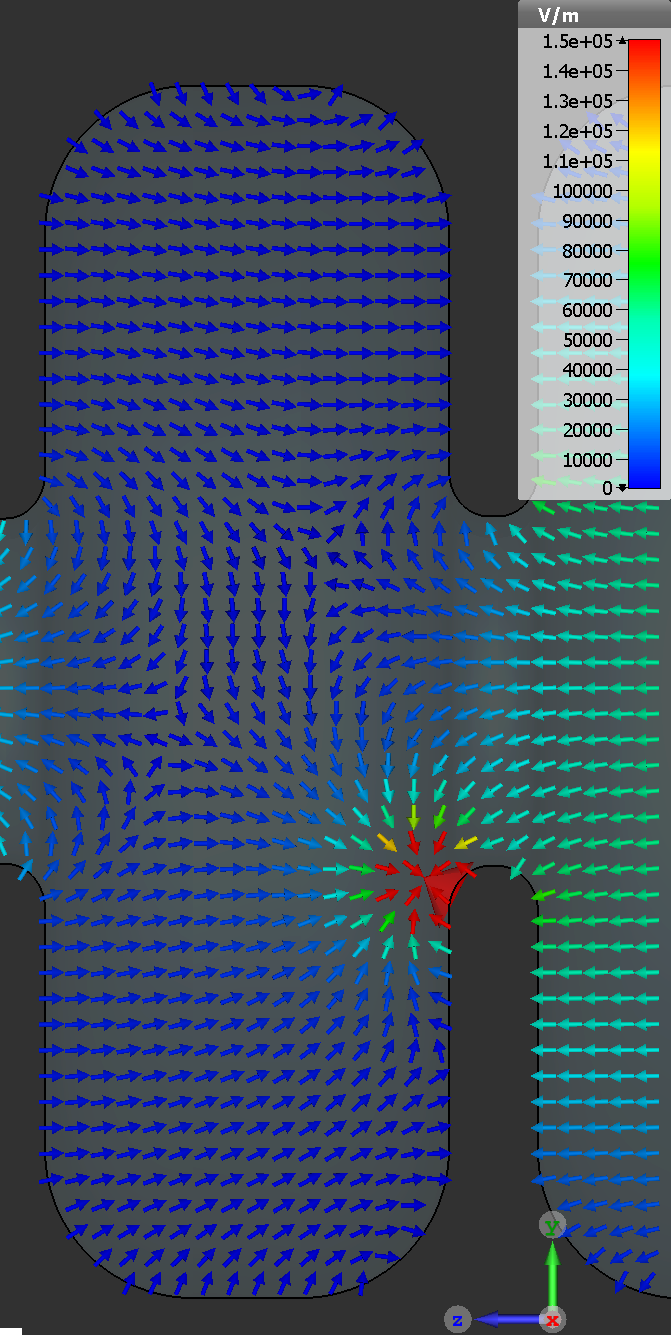}
\caption{Electric-field patterns in a $TM_{010}$ cell. Left: Excited by RF power in the input port as in normal operation. Middle: Real part of the field pattern excited by a sinusoidal current in an antenna. Right: Imaginary part of the field pattern excited by a sinusoidal current in an antenna.}
\label{fig:field_patterns}
\end{figure}

An important aspect to note is that the beam loading effect causes a reduction of the field amplitude in the entire cell, and not only in the immediate vicinity of the antenna. Fig. \ref{fig:field_patterns} shows an example of a simulation showing the electric field pattern resulting from a current in an antenna compared with the electric field pattern in normal operation. One can see that the real part of the electric field (i.e. the component in phase with the antenna current) is very similar to the accelerating mode, which implies that a local emitted current affects the fields in the entire cell.

\subsection{\label{sec:emission_function}Emission Function}
Fowler-Nordheim field emission \cite{Wang1997FieldA} is believed to produce insufficient current to bring about significant beam loading in the entire cell. The electrons from an ionised plasma are a more likely candidate to produce sufficient current. Therefore, it is proposed that the emission function $I = f(E)$ should represent the behaviour of a breakdown in the onset phase, in which the breakdown consumes the most power and therefore RF power coupling is the most critical, rather than the pre-breakdown phase in which the current is produced by field emission. The intermediate breakdown stage is modelled here by the Child-Langmuir law \cite{ChildLangmuir}, which applies if the emission is space-charge-limited, which is the case for very large current densities typical of breakdowns \cite{SLAC_Plasma_Model}. The current density $J$ given by the Child-Langmuir law for a 1D diode with an applied voltage $V$ and length $d$ is:

\begin{equation} \label{eqn:child_langmuir}
    J = \frac{\epsilon_0}{9\pi} \left(\frac{2q_e}{m_e}\right)^{1/2} \frac{V^{3/2}}{d^2}
\end{equation}

A 1D model can be justified if the transverse dimensions of the plasma spot are much larger than the thickness of the emission layer, which would mean that most of the emitted electrons follow 1D trajectories that are initially perpendicular to the metal surface. 

In the context of this study, the total current, as opposed to the current density, is important since it determines the total power absorbed by the breakdown. The Child-Langmuir law describes a linear vacuum diode in which electrons traverse a drift region of length $d$, after which they are absorbed by the anode and cease to affect the system. This is consistent with the use of a short antenna to model the breakdown, whereby the current flow is constrained to the length of the antenna only (see Sec. \ref{sec:antenna_lengths}). With $d = l_{ant}$ and $V = E l_{ant}$, Eq.\eqref{eqn:child_langmuir} can thus be reformulated as:

\begin{equation} \label{eqn:child_langmuir_emission}
    I = k E^{3/2}l_{ant}^{-1/2},
\end{equation}

where $k$ is a proportionality constant comprised of physical constants and effective emitter area. It should be noted that the use of the Child-Langmuir law implies an assumption that the electron velocities are non-relativistic, which might need reviewing. For simplicity, it is assumed that all potential breakdown sites behave identically, though  future iteration of the model could take into account random variations between breakdown sites.

\subsection{\label{sec:analytical_rbd}Analytical Derivation of Breakdown Resistance}
The use of a short antenna to model the breakdown enables an easy way of quantitatively calculating of the impedance of the RF power source as experienced by the breakdown. This can be performed either analytically or with a finite-element electromagnetic simulation.

\begin{figure}[b]
\centering
\includegraphics[width=0.4\textwidth]{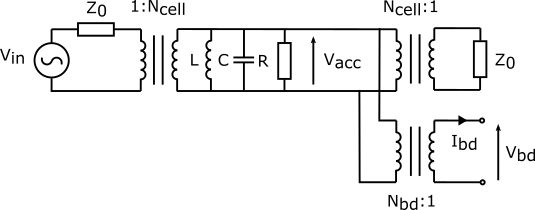}
\caption{Equivalent circuit of a single cell in a travelling-wave configuration.}
\label{fig:single_cell_circuit}
\end{figure}

The analytical calculation of $R_{bd}$ involves a circuit model of the structure including the coupling of power sources and loads. The breakdown is represented by a port coupled to one of the cells via an ideal transformer of ratio $N_{bd}:1$. With such a circuit, $R_{bd}$ is the equivalent resistance of the network as observed through the breakdown port. An example for a single cell is shown in Fig. \ref{fig:single_cell_circuit}.

At the resonant frequency of the cell $f = f_r = 1/2\pi\sqrt{LC}$ \cite{MicrowaveEngineering}, the impedances of the inductor and capacitor cancel each other out, leaving only the real resistance $R$. In most practical travelling-wave structures, the impedance of the structure is also dominated by external coupling rather than resistive losses. The power $P$ flowing through the cell, assuming zero breakdown current, can thus be expressed in terms of circuit parameters as:

\begin{equation} \label{eqn:circuit_model_power}
P = \frac{V_{in}^2}{4Z_0} 
\end{equation}

It can be easily seen that in this circuit, $V_{acc} = N_{bd}V_{bd}$. The value of $N_{bd}$ depends on the location of the breakdown within the cell, which determines the electric field along the antenna for a given accelerating voltage, as given by:

\begin{equation} \label{eqn:bd_transformer}
    N_{bd} = \frac{V_{acc}}{V_{bd}} = \frac{a}{l_{ant}} \frac{E_{acc}}{E_0},
\end{equation}

where $a$ is the length of the cell, $l_{ant}$ is the length of the antenna, $E_0$ is the surface electric field at the location of the antenna and $E_{acc}$ is the accelerating gradient. With some circuit manipulations, the output impedance of the circuit as seen by the breakdown port can also be expressed in terms of circuit parameters:

\begin{equation} \label{eqn:rbd_circuit}
R_{bd} = \frac{Z_0 N_{cell}^2}{2 N_{bd}^2} 
\end{equation}

In the case of no breakdown current, the accelerating voltage can be expressed in terms of the input voltage as $V_{acc} = N_{cell}V_{in}/2$. With this,  Eq. \eqref{eqn:circuit_model_power}, and the relationship between power and group velocity \cite{Wangler2008}, given by:

\begin{equation}
    P = \frac{v_g a}{2 \pi f} \frac{Q}{R} E_{acc}^2,
\end{equation}

where $P$ is the incident power from the source, $f$ is the operating frequency, $v_g$ is the group velocity, $a$ is the length of the cell, $R$ is the shunt impedance, and $Q$ is the quality factor of the structure, the relationship between input power and accelerating voltage can be derived as:

\begin{equation} \label{eqn:rbd_z0}
    \frac{V_{acc}^2}{P} = \frac{2 \pi a f}{v_g} \frac{R}{Q} = N_{cell}^2 Z_0.
\end{equation}

$R_{bd}$ can thus be expressed entirely in terms of quantities obtainable from an eigenmode solution of the electromagnetic wave equation in the cell geometry as:

\begin{equation} \label{eqn:analytical_rbd_single_freq}
    R_{bd} = \frac{\pi f_r l_{ant}^2}{v_g a} \frac{R}{Q} \left(\frac{E_0}{E_{acc}}\right)^2.
\end{equation}

\begin{figure}[b]
\centering
\includegraphics[width=0.45\textwidth]{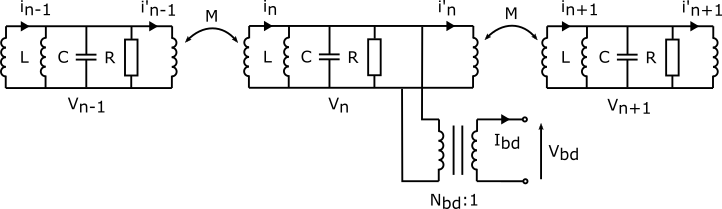}
\caption{Equivalent circuit of an infinite array of identical coupled resonant cells, with a port in one cell representing the coupling to a breakdown.}
\label{fig:coupled_cells_rbd_circuit}
\end{figure}

If transient behaviour needs to be taken into account (see sec. \ref{sec:broadband_correction}), the broadband behaviour of the structure needs to be modelled. Fig.~\ref{fig:coupled_cells_rbd_circuit} shows a circuit model of a structure with an infinite number of identical cells coupled with mutual inductances $M$ instead of ideal transformers. One of the cells is coupled to a breakdown port via an ideal transformer as before. The array is assumed to have a travelling wave flowing through it, with a phase advance $ka$ from cell to cell. Thus, the $n$\textsuperscript{th} cell is assumed to have voltage and current values $v_n$ and $i_n$ given by:

\begin{eqnarray} \label{eqn:circuit_model_vn_in}
    v_n &&= v_0 e^{jkan}\\
    i_n &&= i_0 e^{jkan}
\end{eqnarray}

Using circuit theory, one can obtain an expression for the dependence of currents and voltages in adjacent cells:

\begin{eqnarray} \label{eqn:circuit_model_kirchoff}
    &&v_{n+1} = j 2 \pi f M i'_n \nonumber\\&&= j 2 \pi f M \left(i_n - \frac{v_n}{j 2 \pi f  L} - j 2 \pi f  C v_n - \frac{v_n}{R} \right), 
\end{eqnarray}

as well as the characteristic impedance $Z_0$ of this transmission line:

\begin{equation} \label{eqn:z0_definition}
    Z_0 = \frac{v_n}{i_n} = -j 2 \pi f  M e^{jka}.
\end{equation}

Eqs. \eqref{eqn:circuit_model_vn_in} and \eqref{eqn:circuit_model_kirchoff} allow the dispersion relation of this transmission line to be expressed as:

\begin{eqnarray} \label{eqn:circuit_model_kirchoff_3}
    &&e^{jka} + e^{-jka} = 2 \cos(ka) \nonumber\\ &&= - j 2 \pi f M \left(\frac{1}{j 2 \pi f L} + j 2 \pi f C + \frac{1}{R} \right).
\end{eqnarray}

By substituting $0$ and $\pi$ for the phase advance $ka$ into the dispersion relation \eqref{eqn:circuit_model_kirchoff_3} and defining the resulting values of $f$ as $f_0$ and $f_{\pi}$ respectively, as well as using the definitions of resonance frequency $f_r = 1/2 \pi \sqrt{LC}$ and quality factor $Q = R\sqrt{L/C}$, an expression for $M$ may also be obtained:

\begin{equation} \label{eqn:M}
    M = \frac{2}{\pi} \frac{R}{Q} \frac{f_0}{f_{\pi}^2 - f_0^2},
\end{equation}

leading to an expression of the dispersion relation without circuit values:

\begin{equation} \label{eqn:dispersion_relation}
    f^2 = f_r^2 - \frac{f_{\pi}^2 - f_0^2}{2} \cos(ka).
\end{equation}

Using \eqref{eqn:M} and \eqref{eqn:dispersion_relation}, one can also eliminate the parameters $M$ and $ka$ from the expression for characteristic impedance \eqref{eqn:z0_definition}, yielding:

\begin{equation} \label{eqn:z0_eigen}
    \Re(Z_0(f)) = \frac{R}{Q} \frac{4f f_r}{f_{\pi}^2 - f_0^2} \left(1-\left(2\,\frac{f_r^2 - f^2}{f_{\pi}^2 - f_0^2}\right)^2\right)^\frac{1}{2}.
\end{equation}

The $R_{bd}$ value in this case can be determined from the characteristic impedance $Z_0$ of the structure:

\begin{eqnarray} \label{eqn:analytical_rbd}
    &&R_{bd}(f) = \frac{\Re(Z_0(f))}{2N_{bd}^2}= \nonumber\\
 && \frac{2f f_r}{f_{\pi}^2 - f_0^2} \left(1-\left(2\frac{f_r^2 - f^2}{f_{\pi}^2 - f_0^2}\right)^2\right)^\frac{1}{2} \nonumber\\ &&\times \frac{R}{Q} \left(\frac{l_{ant}}{a}\right)^2 \left(\frac{E_0}{E_{acc}}\right)^2,
\end{eqnarray}

where $f_0$, $f_r$, $f_\pi$ are frequencies corresponding to a cell-to-cell phase advance values of zero, $\pi/2$, and $\pi$ respectively. Standing-wave structures will not be considered specifically in this paper, though it is expected that an analogous approach, using a circuit model with a finite number of cells coupled to a transmission line, should give appropriate results.

\subsection{\label{sec:antenna_lengths}Antenna Lengths}
Both of the analytical models presented in Sec. \ref{sec:analytical_rbd} exhibit a quadratic dependence of $R_{bd}$ on the antenna length, i.e.: $R_{bd} \propto l_{ant}^2$. This resembles the behaviour of the radiation resistance of a small Hertzian dipole whose length $\l_{ant}$ is much smaller than the free-space wavelength $\lambda$ \cite{AntennaTheory}:

\begin{equation} \label{eqn:Hertz_dipole}
    R_{rad} = \frac{2\pi}{3} \zeta_0 \left( \frac{l_{ant}}{\lambda} \right)^2,
\end{equation}

where $R_{rad}$ is the radiation resistance, and $\zeta_0$ is the impedance of free space. The antenna length is the only parameter in Eqs. \eqref{eqn:analytical_rbd_single_freq} and \eqref{eqn:analytical_rbd} not already defined by the geometry of the structure. It is not immediately clear what value of antenna length should be chosen to best model a breakdown. It may not be practical in general to use a particular antenna length in all cases, due to limitations imposed by the geometry in question. The issue of choosing a value for antenna length can be circumvented by introducing a quantity $R_0$ that is independent of $l_{ant}$, such that:

\begin{equation} \label{eqn:r0_definition}
    R_{bd} = R_0 l_{ant}^2
\end{equation}

As per the definition of $R_{bd}$ in Eq. \eqref{eqn:bd_impedance_approx}, 

\begin{equation} \label{eqn:bd_resistance_normalised}
\begin{aligned}
    \Re (\Delta V) &= I R_{bd} = I R_0 l_{ant}^2\\
    \Re (\Delta E) &= \frac{\Re (\Delta V)}{l_{ant}} = Il_{ant} R_0
    \end{aligned}
\end{equation}

Thus, if the breakdown loading plot (such as the one in Fig. \ref{fig:example_impedance}), is modified such that the axes are the electric field $E$ and the product $Il_{ant}$, the plot and equilibrium solution become independent of antenna length. To suit, the emission function can also be expressed as:

\begin{equation} \label{eqn:bd_emission_characteristic_normalised}
    Il_{ant} = k' E^{3/2},
\end{equation}

where $k = k' l_{ant}^{-1/2}$. The value of $k'$ can be chosen independently of the antenna length. Eq. \eqref{eqn:e_loading_solution} can thus be expressed as the solution to the equation:

\begin{equation}
\label{eqn:eloaded_solution}
    k' R_0 E^{*3/2} + E^* - E_0 = 0,
\end{equation}

which is independent of antenna length. As the electrons emitted during a breakdown continue to absorb energy from the local electric field until they collide with the wall of the structure or escape, the amount of energy they absorb depends on the length of their trajectories. The fact that the quantity $Il_{ant}$ depends on antenna length reflects this behaviour, and picking $l_{ant}$ to be a value representative of the mean distance travelled by the emitted particles appears reasonable. It is difficult to determine the exact physical length of a breakdown during the onset phase in every possible breakdown location in a given geometry without a detailed particle tracking simulation study but it is most reasonable to assume it scales approximately linearly with the size of the cell. If this is the case, the same $k'$ value should be used for structures with the same operating frequency, and thus similar dimensions. It may be the case for some geometries that $l_{ant}$ is independent of frequency. For example, the electron trajectory could be constrained by physical obstacles, such as a narrow waveguide or the vanes of a radiofrequency quadrupole (RFQ). Such effects are briefly considered in Sec. \ref{sec:thin_gaps}, but require further work.

\subsection{\label{sec:broadband_correction}Broadband Effects}
As breakdown is by nature a transient phenomenon, the transient behaviour of the system needs to be considered. Introducing transient behaviour requires information about $R_{bd}$ over the full bandwidth of the structure. In contrast, the previous case of calculating $E^*$ using $R_{bd}$ at a single fixed frequency implies a quasi-steady-state scenario in which the local stored energy plays a relatively minor role, like in high-group-velocity structures.

It was decided to use a scalar impedance-like quantity that is representative of the transient behaviour of the structure. Following an impulse excitation current $i(t) = Q\delta(t)$, $Q$ being the emitted charge and $\delta(t)$ begin the Dirac delta function, the transient voltage across the antenna is given by $Qr_{bd}(t)$, where $r_{bd}(t)$ is the inverse Fourier transform of $R_{bd}(f)$. The total energy $W$ dissipated by the breakdown (approximated by a linear load resistance $R_l$) will thus be given by:

\begin{equation} \label{eqn:broadband_derviation}
    W = \frac{Q^2}{R_l} \int_{-\infty}^{\infty} r_{bd}^2(t) dt = \frac{Q^2}{R_l} \int_{-\infty}^{\infty} R_{bd}^2(f) df.
\end{equation}

A distinction should be made between the energy that results in the heating of the emission site and the total energy absorbed by the breakdown event. Much of the energy consumed by a breakdown results in the acceleration of electrons to high energies which then either escape the structure or impact a site relatively far from the breakdown site, neither of which contribute to the heating of the breakdown site. If it is the latter that is important, then $R_l$ should represent the load resistance of the breakdown site rather than the entire breakdown. The quantity $L$ was defined in order to describe the influence of the structure geometry on the heating of the emission site:

\begin{equation} \label{eqn:square_integral_rbd}
    L = \sqrt{\int_{0}^{\infty}R_{bd}(f)^2 df}.
\end{equation}

While the value of $R_l$ is unknown, it can be absorbed by the proportionality constant $k$ of the emission function, as long as all candidate breakdown sites have similar properties. The square root of the integral is taken to obtain a quantity that scaled linearly with $R_{bd}$. Analogously to $R_0$, a quantity $L_0$ that is independent of antenna length can also be introduced such that $L = L_0 l_{ant}^2$, which cancels out the $l_{ant}^2$ term seen in Eq.\eqref{eqn:analytical_rbd}.

\subsection{\label{sec:numerical_rbd}Numerical Calculation of Breakdown Resistance}
The concept of using a monopole antenna to model the electromagnetic coupling of a breakdown can be implemented directly in a numerical simulation, which allows $R_{bd}$ to be determined for arbitrary geometries and frequencies. In this study, simulations were performed using CST Microwave Studio \cite{CST}. To model the breakdown, a Discrete Current Port was added to the geometry, with one end point located on the surface of the structure at the location of the prospective breakdown site, while the other end point was located a distance $l_{ant}$ away, with the antenna normal to the surface. A frequency-domain simulation was run with the antenna used as a constant-amplitude current source. The $R_{bd}$ value for this location was calculated using Eq.\eqref{eqn:bd_impedance_approx}. Since a current port in Microwave Studio models an ideal current source, the impedance of the port itself is infinite, meaning that no power is absorbed by a port not currently set as the excitation source. Thus, several such ports could be incorporated into the model simultaneously, allowing the $R_{bd}$ to be calculated at multiple locations using the same mesh. The relation in Eq.~\eqref{eqn:r0_definition} can be used to scale results obtained with different $l_{ant}$ values, which means that $l_{ant}$ can be chosen for optimal meshing performance in a specific simulation setup.

The $R_{bd}$ value can also be obtained from a finite-element simulation by using the $Z$-matrix \cite{MicrowaveEngineering} of the simulated geometry:

\begin{equation} \label{eqn:z_param}
    R_{bd} = \Re \left(Z_{a,a} - \frac{Z_{a,i} Z_{i,a}}{Z_0 + Z_{i,i}} \right),
\end{equation}

where $Z_0$ is series impedance of the RF power source, and the subscripts $i$ and $a$ indicate terms of the $Z$-matrix corresponding to the RF input port and the antenna respectively. In CST Microwave Studio, this requires the antennas to be set as S-parameter ports with a finite characteristic impedance. A sufficiently large value of $Z_0$ allows several antennas to be incorporated into the same simulation with negligible loading effects, as with the current sources.

\section{\label{sec:c3}Comparison with $S_c$ for $TM_{010}$ structures}
There is ample experimental data that $S_c$ accurately predicts the breakdown performance for RF structures operating in the $TM_{010}$ mode, so it can be used as a benchmark for the new antenna-based model in this limited parameter space. Thus, the predictions of the $E^*$ model should be in line with those of $S_c$ for such structures. To investigate this, a series of hypothetical $TM_{010}$ travelling-wave accelerating structure geometries was simulated to compare $E^*$ and $S_c$ values inside them.

\begin{table}[]
\caption{\label{tab:c3_params}Parameters common to all of the C3 structures simulated.}
\centering
\begin{ruledtabular}
\begin{tabular}{lcc}
Parameter            & Value & Unit \\ \hline
Operating frequency  & 12    & GHz  \\
Phase advance        & 120   &  \degree \\
Phase velocity / c   & 1     &   -   \\
Periodicity          & 8.333 & mm   \\
Iris thickness     & 2     & mm   \\
Iris rounding radius & 1     & mm   \\
Cell rounding radius & 2     & mm  
\end{tabular}
\end{ruledtabular}
\end{table}

\begin{figure}[htbp]
\centering
\includegraphics[width=0.22\textwidth]{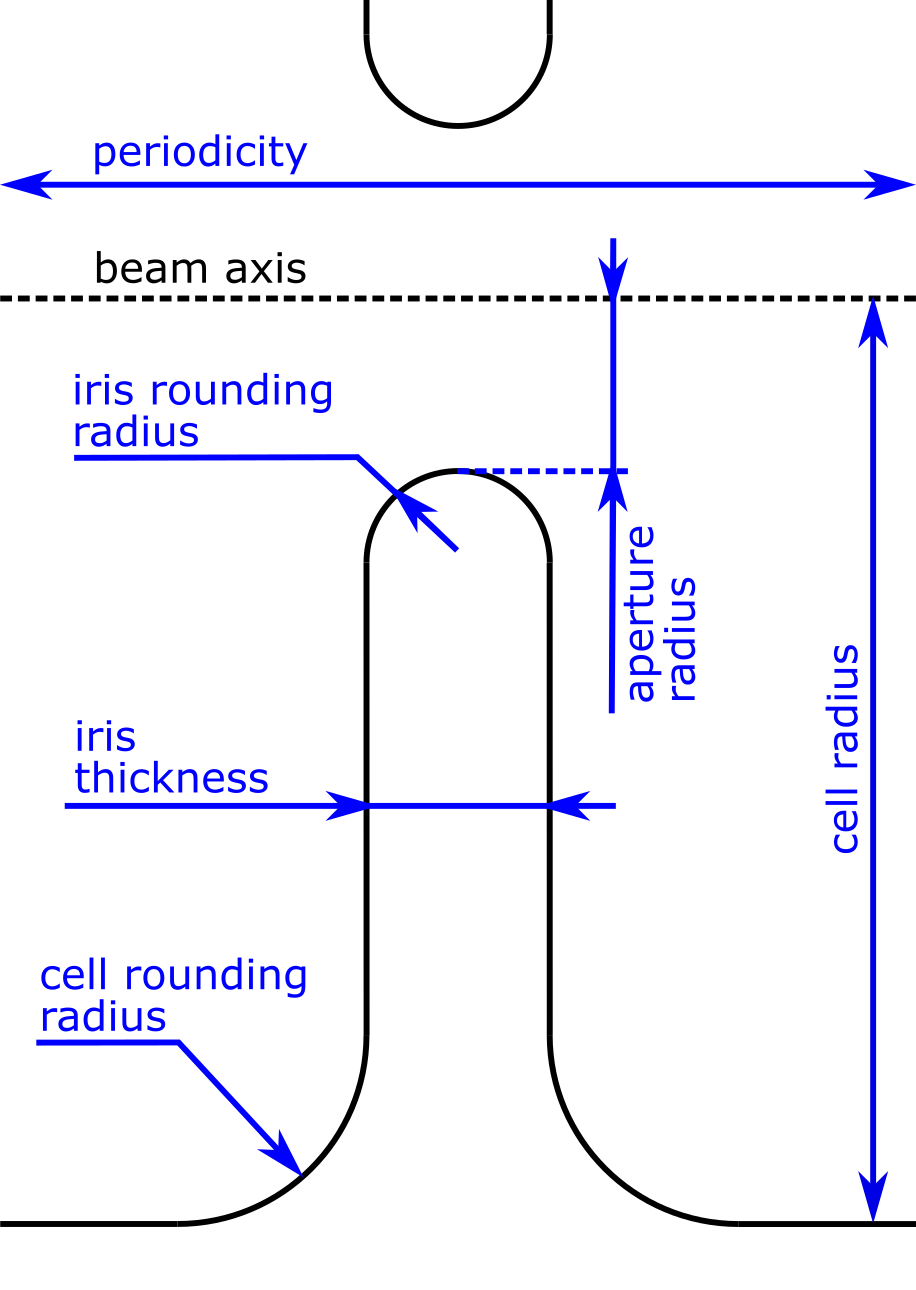}
\includegraphics[width=0.22\textwidth]{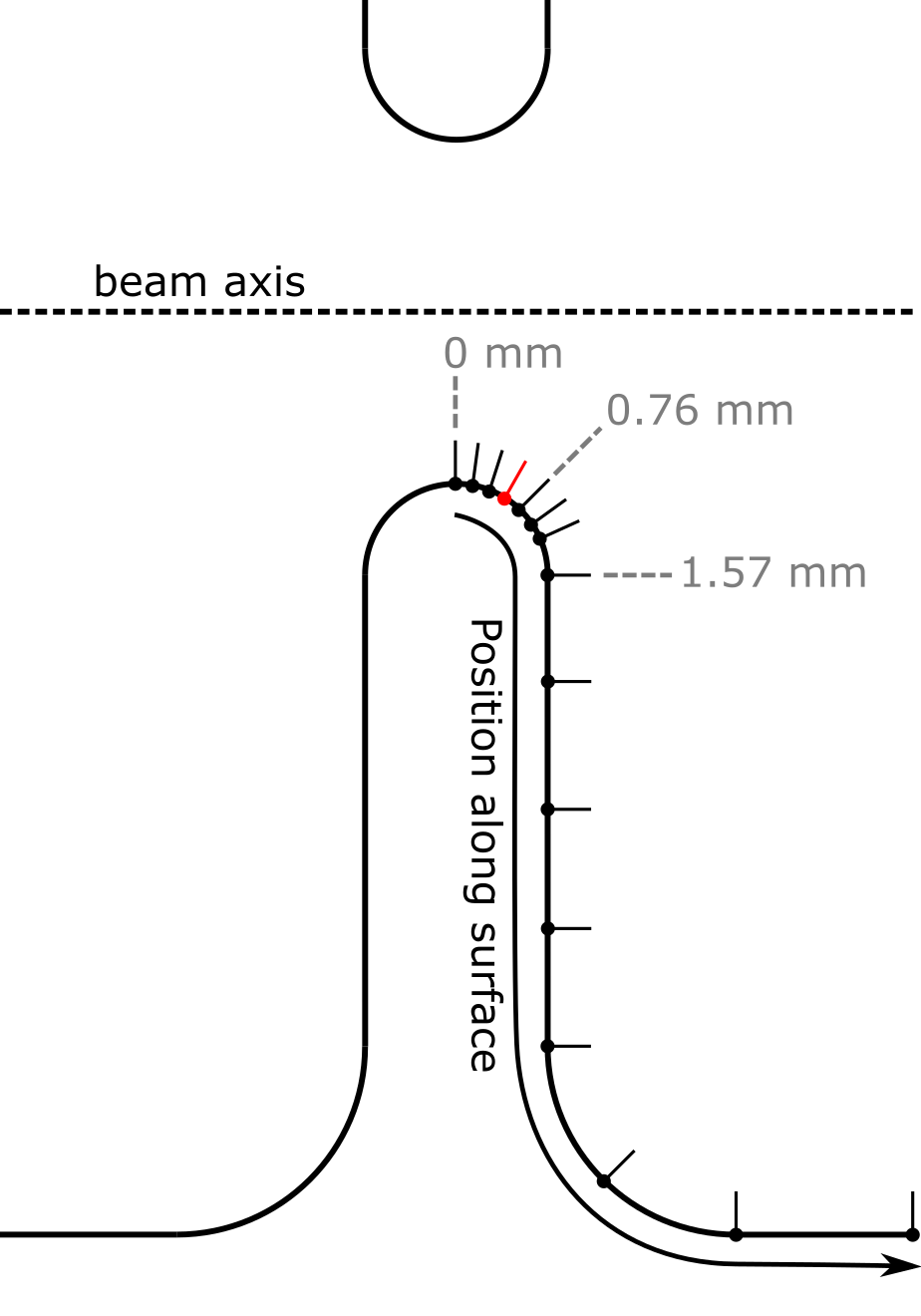}
\caption{Left: dimensions of a regular cell. Right: placement of antennas for numerical $R_{bd}$ calculations, with markings showing the distances from the iris apex in millimetres. Antenna lengths not to scale.}
\label{fig:c3_dimensions}
\end{figure}

Due to the availability of a relatively large body of experimental data on CLIC-like structures, the parameters of the simulated structures were made similar to CLIC structures so that a meaningful comparison could be made. Hence, a travelling wave design with a range of aperture radii varying from 2.5 mm to 6 mm was simulated. The outer radius of the cells was adjusted for each aperture size to maintain a 120\degree phase advance from cell to cell at 12 GHz. The dimensions of each regular cell are shown in Fig~\ref{fig:c3_dimensions}, with values given in Tab.~\ref{tab:c3_params}.

\begin{table}[]
\caption{\label{tab:c3_matching_params}Geometrical parameters of each of the simulated C3 structures.}
\centering
\begin{ruledtabular}
\begin{tabular}{cccc}
\multicolumn{1}{l}{$r_{iris}$ {[}mm{]}} & \multicolumn{1}{l}{$r_{cell}$ {[}mm{]}} & \multicolumn{1}{l}{$r_{iris, match}$ {[}mm{]}} & \multicolumn{1}{l}{$r_{cell, match}$ {[}mm{]}} \\ \hline
2.5                                       & 10.083                                    & 4.803                                              & 10.438                                            \\
3                                         & 10.211                                    & 5.316                                              & 10.608                                            \\
3.5                                       & 10.373                                    & 5.765                                              & 10.785                                            \\
4                                         & 10.563                                    & 6.194                                              & 10.975                                            \\
5                                         & 11.026                                    & 6.962                                              & 11.381                                            \\
6                                         & 11.584                                    & 7.660                                              & 11.817                                           
\end{tabular}
\end{ruledtabular}
\end{table}

Since calculating $E^*$ requires a consideration of power flow, a calculation of the eigenmode field patterns was insufficient as it would not be useful to simulate the resonant cells alone without specifying how RF power was coupled into them. Thus, the simulation also had to include matched waveguide ports at each end of the structure, coupled to circular waveguides of 12 mm radius. To  ensure an impedance match between the structure and the circular waveguides, matching cells were incorporated between the regular cells and the input and output waveguides. The radius of the matching cells and the outermost irises was chosen independently of the parameters of the regular cells to minimise the reflected power at 12 GHz, with the optimised parameter values shown in Table  \ref{tab:c3_matching_params}.The structure featured three identical regular cells, with the middle cell containing a number of antennas on its surface. The antennas were placed in a line from the apex of one of the irises bordering the centre cell, to the middle of the wall of the centre cell, as illustrated in Fig. \ref{fig:c3_dimensions}.

\begin{figure}[htbp]
\centering
\includegraphics[width=0.5\textwidth]{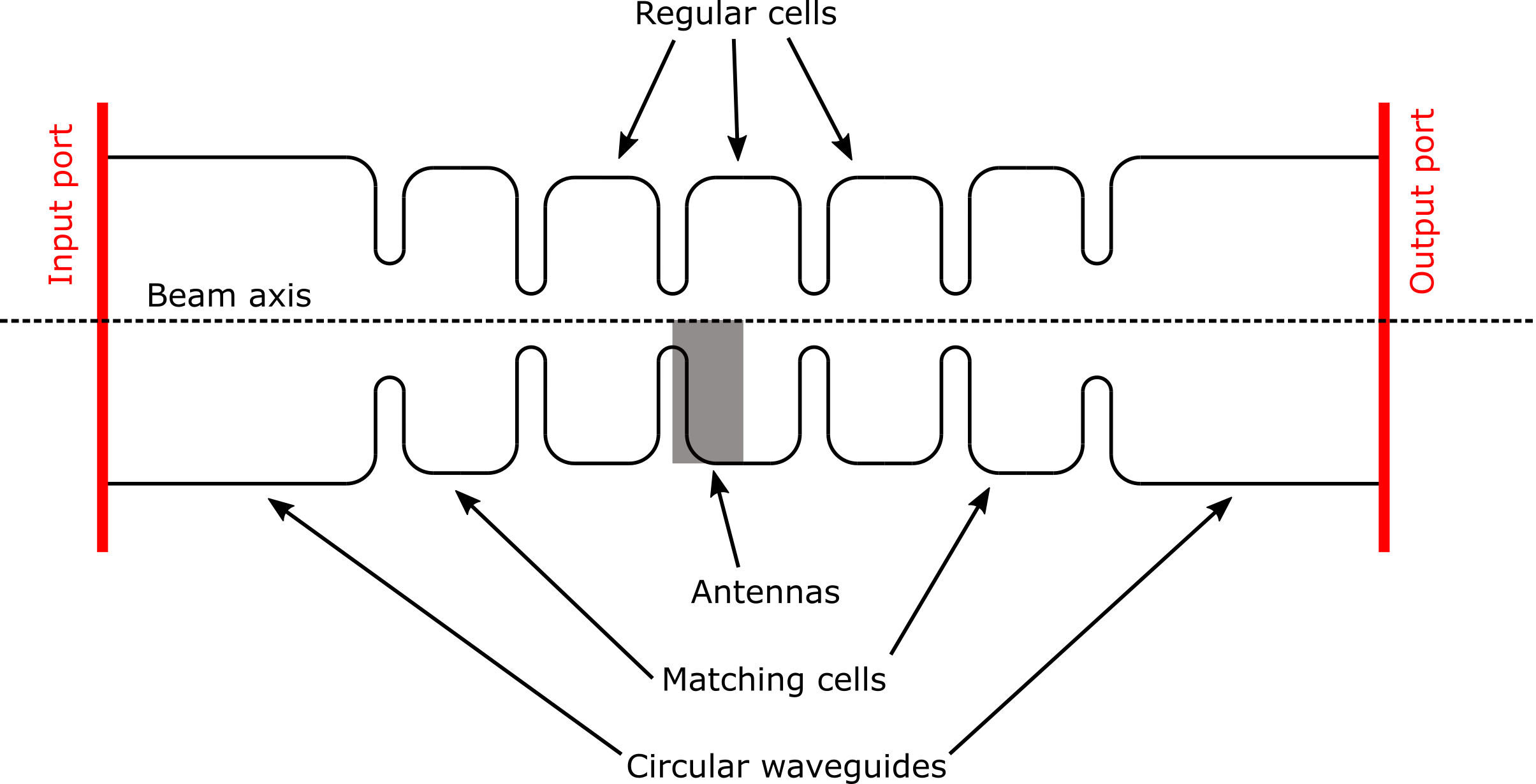}
\caption{Diagram of the full structure simulated. The shaded area represents the region in which antennas were placed for calculating $R_{bd}$.}
\label{fig:c3_full_structure_diagram}
\end{figure}

Three regular cells were used to avoid any distortions of the results caused by evanescent fields from the matching cells bleeding into the cell containing the antennas. This helped make the conditions in the centre cell more similar to those within an infinitely long constant-impedance structure.

A diagram of the full geometry simulated is shown in Fig.~\ref{fig:c3_full_structure_diagram}. The simulations were performed using CST Microwave Studio as described in Sec. \ref{sec:numerical_rbd} to calculate $E_0$ and $R_{bd}$ values at locations inside the structure.

\begin{figure}[htbp]
\centering
\includegraphics[width=0.5\textwidth]{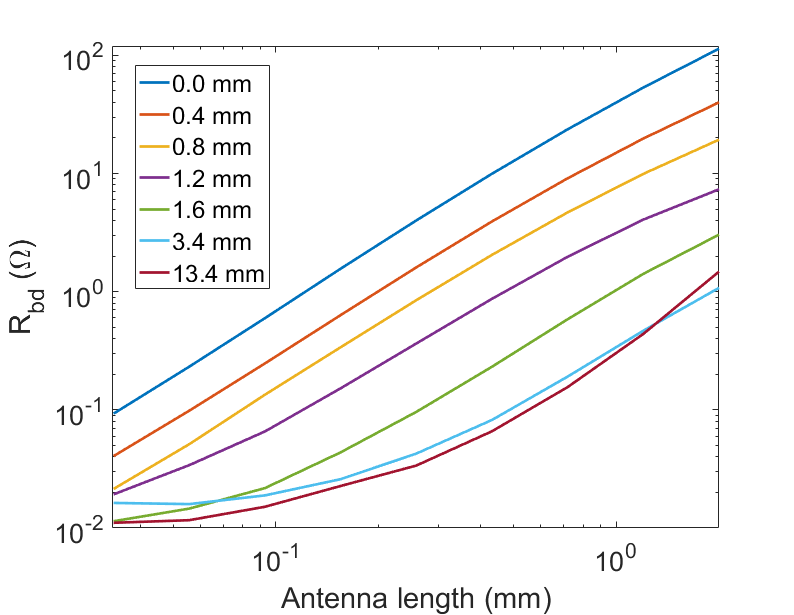}
\caption{Calculated $R_{bd}$ in $\Omega$ vs. antenna length in mm for antennas in different locations in a C3 structures of 4 mm aperture radius. The distance of the antenna from the iris apex, as shown in Fig.~\ref{fig:c3_dimensions}, is indicated in the legend.}
\label{fig:rbd_vs_ant_l_mesh_comparison}
\end{figure}

\begin{figure}[htbp]
\centering
\includegraphics[width=0.5\textwidth]{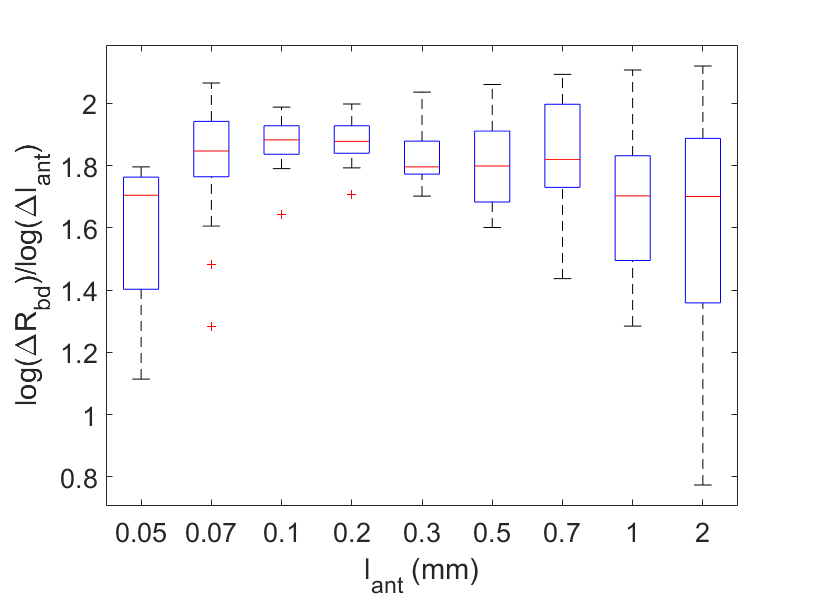}
\caption{Distribution of $log(\Delta R_{bd}) / log(\Delta l_{ant})$, i.e. the slope on a logarithmic plot of $R_{bd}$ vs. $l_{ant}$ for different antenna lengths and aperture sizes.}
\label{fig:rbd_vs_lant_statistics}
\end{figure}

Simulations were performed with various antenna lengths varied to find the optimal value. Fig. \ref{fig:rbd_vs_ant_l_mesh_comparison} shows the dependence of $R_{bd}$ on antenna length for different antenna locations for a 4 mm iris radius. For values of $l_{ant}$ between 0.1 and 1 mm, the relation $R_{bd} \propto l_{ant}^2$ held for most of the antennas. For most of the antennas, the quadratic relationship did not hold outside of this range. It is believed that this was due to numerical noise for very short antennas, and the assumption that $E$ is constant over the length of the antenna no longer being valid for very long antennas. Since an antenna length of 0.1 mm resulted in the closest adherence to the expected $R_{bd} \propto l_{ant}^2$, as illustrated in Fig. \ref{fig:rbd_vs_lant_statistics}, this was the length value at which the final $R_{bd}$ values were obtained.

\begin{figure}[htbp]
\centering
\includegraphics[width=0.5\textwidth]{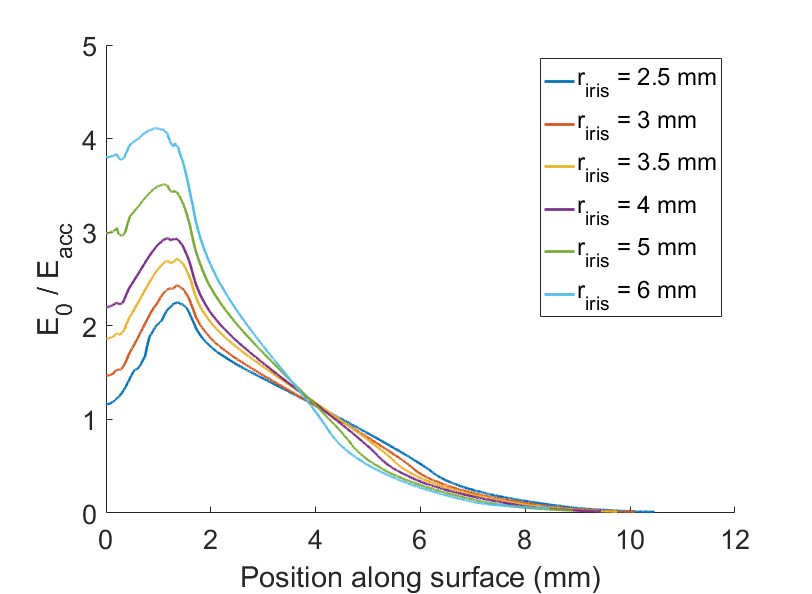}
\caption{Unloaded surface electric field at unit accelerating gradient vs. position along the surface in mm. The position is defined by the arrow shown in Fig.~\ref{fig:c3_dimensions}, where 0 is the iris apex. Different coloured plots represent C3 structures of different aperture sizes. The discontinuities near the 0.5 mm and 1.5 mm marks on some of the plots are likely the result of the rounded geometry of the iris being discretised as a tetrahedral mesh in the finite-element simulation.}
\label{fig:c3_e0_profile}
\end{figure}

\begin{figure}[htbp]
\centering
\includegraphics[width=0.5\textwidth]{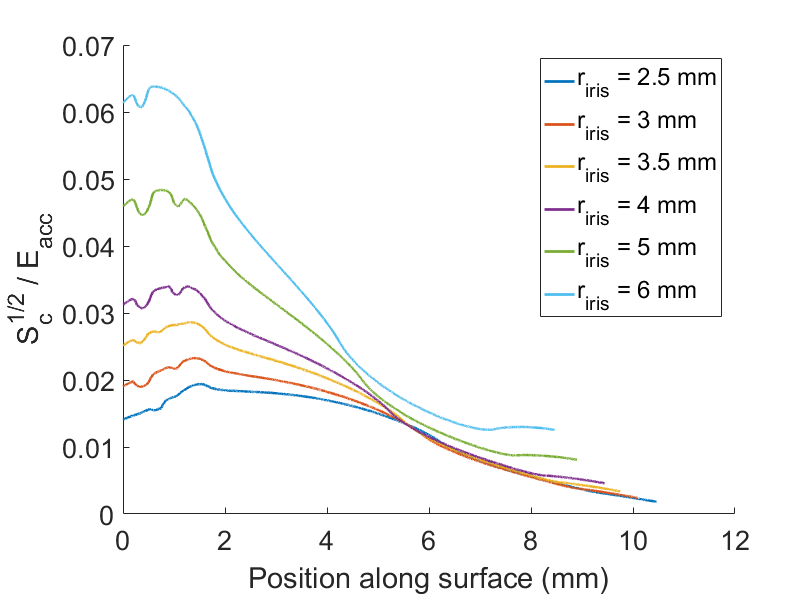}
\caption{$\sqrt{S_c}$ at unit accelerating gradient vs. position along the surface in mm. The position is defined by the arrow shown in Fig.~\ref{fig:c3_dimensions}, where 0 is the iris apex. Different coloured plots represent C3 structures of different aperture sizes. The discontinuities near the 0.5 mm and 1.5 mm marks on some of the plots are likely the result of the rounded geometry of the iris being discretised as a tetrahedral mesh in the finite-element simulation.}
\label{fig:c3_sc_profile}
\end{figure}

\begin{figure}[htbp]
\centering
\includegraphics[width=0.5\textwidth]{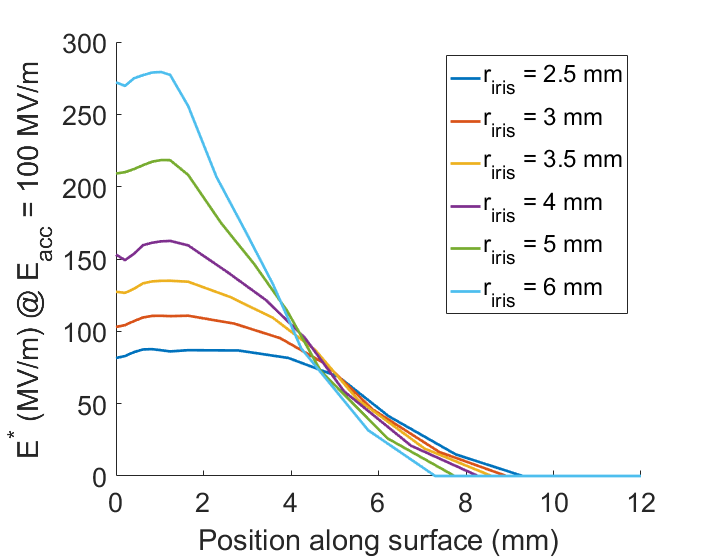}
\caption{$E^*$ at an accelerating gradient of 100 MV/m vs. position along the surface in mm. The position is defined by the arrow shown in Fig.~\ref{fig:c3_dimensions}, where 0 is the iris apex. Different coloured plots represent C3 structures of different aperture sizes. The discontinuities near the 0.5 mm and 1.5 mm marks on some of the plots are likely the result of the rounded geometry of the iris being discretised as a tetrahedral mesh in the finite-element simulation.}
\label{fig:c3_transient_eloaded_profile}
\end{figure}

A value of $k' = 2.30\times10^{-17}$ Am\textsuperscript{2}V\textsuperscript{-3/2} was chosen along with a threshold $E^*$ of 105 MV/m, which produced results consistent with experimental data (see Sec.\ref{sec:t24_vs_crab_results}).

$E^*$ and $S_c$ were compared in two ways: the expected breakdown location, i.e. the physical location in the geometry where the maximum value of the quantity is found, and the prediction of the maximum achievable accelerating gradient. The breakdown location is compared in Figs. \ref{fig:c3_e0_profile}, \ref{fig:c3_sc_profile}, \ref{fig:c3_transient_eloaded_profile}, which show $E_0$, $S_c$, and $E^*$ vs. position on the surface of the middle cell respectively. $E_0$ and $S_c$ have been normalised to accelerating gradient, whereas $E^*$, not scaling linearly with gradient, has been plotted for an accelerating gradient of 100 MV/m. Both $S_c$ and $E_0$ have peaks between the 1 mm and 2 mm marks, which corresponds roughly to experimentally observed breakdown locations, i.e. a short distance from the iris apex at the location of maximum $E_0$.

\begin{figure}[htbp]
\centering
\includegraphics[width=0.5\textwidth]{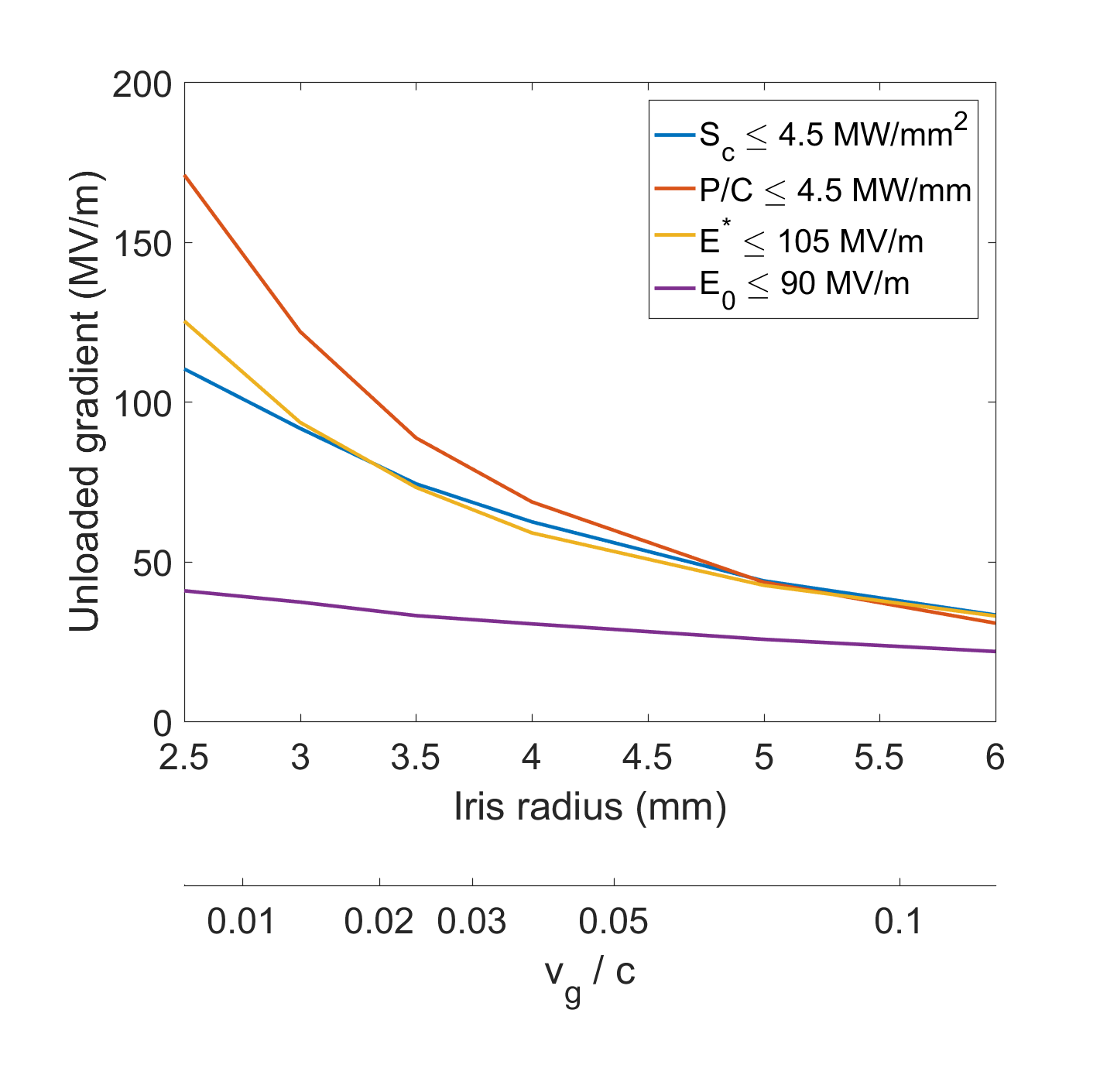}
\caption{Maximum accelerating gradient without beam loading vs. aperture size and group velocity as a fraction of $c$.}
\label{fig:c3_max_gradient_transient}
\end{figure}

Predictions of the maximum gradient according to various breakdown quantities are represented in Fig. \ref{fig:c3_max_gradient_transient}. For each quantity, a certain maximum permissible value was chosen. The values $S_c = $ 4.52 MW/mm\textsuperscript{2}, $P/C = $ 4.8 MW/mm, and $E_0 = $ 90 MV/m (the Kilpatrick limit at 12 GHz) are used as limiting quantities along with $E^* = $ 105 MV/m (see Sec. \ref{sec:t24_vs_crab_results}). Similar behaviour can be seen for all three quantities for iris radii of 3.5 mm and above. The maximum gradient prediction of $P/C$ diverges for smaller iris radii, and therefore smaller group velocities. This is likely due to the contribution of energy stored in the structure, which is not considered by the $P/C$ quantity, yet becomes more significant as the group velocity decreases.

\section{\label{sec:comparison_with_experiments}Comparison with Experimental Results}
$E^*$ was calculated for two experimentally-tested X-band structure geometries to allow a comparison to experimental results, and to check if the method shows consistent behaviour and its usefulness a predictor of breakdown performance. Both of the structures discussed in this section are, like the idealised structures discussed in Sec. \ref{sec:c3}, 12 GHz travelling-wave structures. The same general approach as before to the numerical calculation of $E^*$ was used, with $k' = 2.30\times10^{-17}$ Am\textsuperscript{2}V\textsuperscript{-3/2}, the same value as used in Sec. \ref{sec:c3}. The $R_{bd}$ values were obtained from simulations performed with $l_{ant} =$ 0.5 mm, followed by scaling to $l_{ant} =$ 0.1 mm to make meaningful comparisons with the results from Sec. \ref{sec:c3}.

In should be noted that in the $E^*$ model, conditioning would correspond to the increase of the $E^*$ value at which breakdown occurs. To model this, the threshold $E^*$ value should therefore be increased with conditioning progress. However, experiments generally show the accelerating gradient tending to some limit as conditioning progresses. The experimentally-tested structures discussed in this section have been fully conditioned and the limiting gradient values will be compared, to ensure that the comparison is made between structures in similar states of conditioning.

\subsection{\label{sec:crab_results}CLIC Crab Cavity}
One of the structures considered here is the CLIC Crab Cavity prototype discussed in Sec. \ref{sec:review}, in order to determine whether or not the $E^*$ model resolves the discrepancy in location of breakdowns between the predictions of $S_c$ and experimental results.

\begin{figure}[htbp]
\centering
\includegraphics[width=0.5\textwidth]{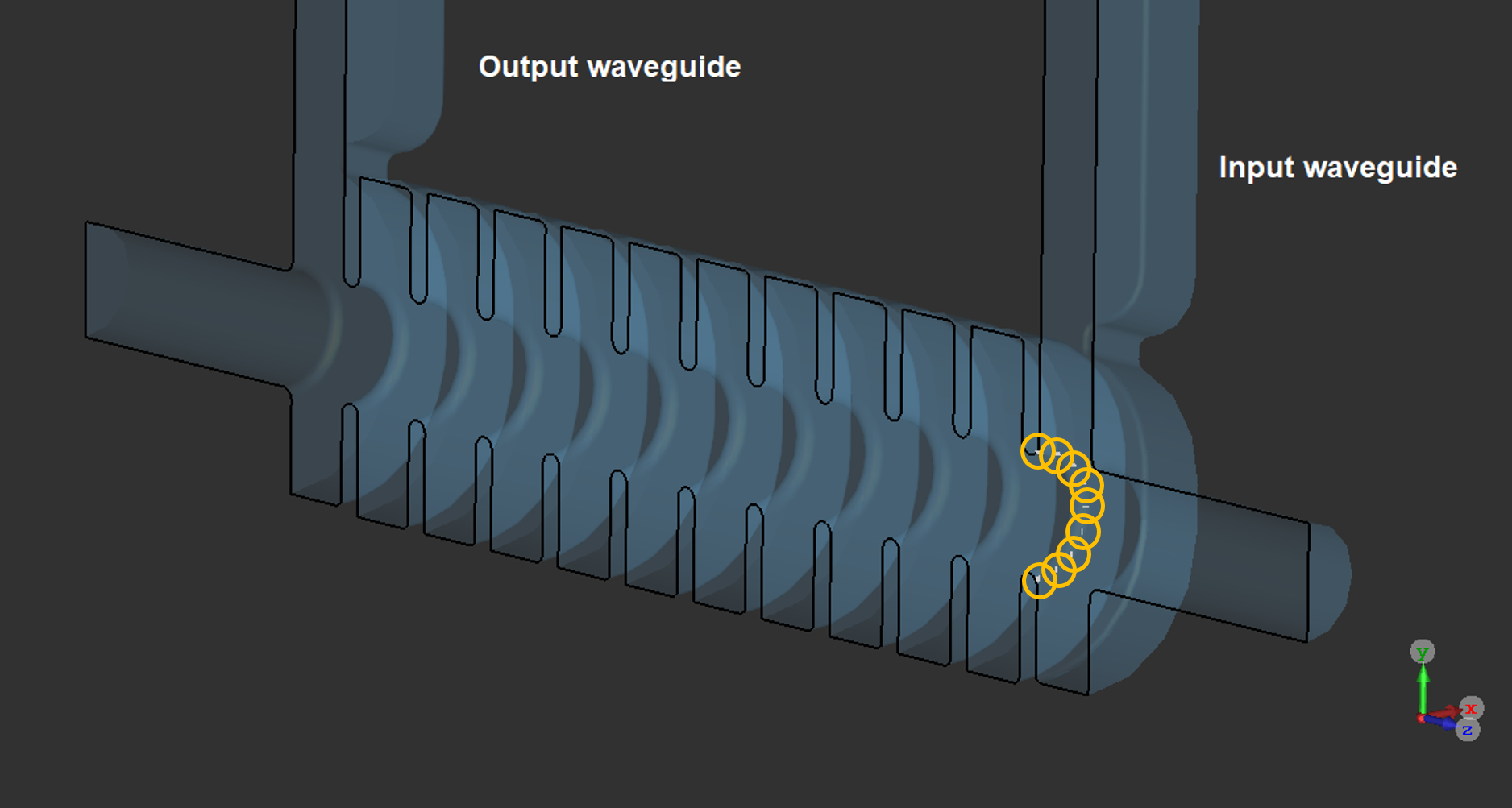}
\caption{A cutaway view of the internal volume of the CLIC Crab Cavity with the locations of antennas in the coupling cell shown as white lines with yellow circles around them.}
\label{fig:crab_antenna_setup}
\end{figure}

Antennas were placed at various azimuths around the iris of the input coupling cell and first regular cell. When this structure was tested in XBox 2, most of the breakdowns were found to occur in these two cells, with a very small proportion of the total breakdown count located in the remaining cells. This is typical of constant-impedance travelling-wave structures, in which the power and field levels gradually decrease along the structure due to Ohmic losses. Nine antennas were used for each of the two cells, forming a 180\degree arc in each case. The antennas were placed slightly away from the iris apex at the location of maximum surface electric field, in order to characterise the limiting field value of the structure. The location of the antennas in the coupling cell is illustrated in Fig.~\ref{fig:crab_antenna_setup}.

\begin{figure}[htbp]
\centering
\includegraphics[width=0.4\textwidth]{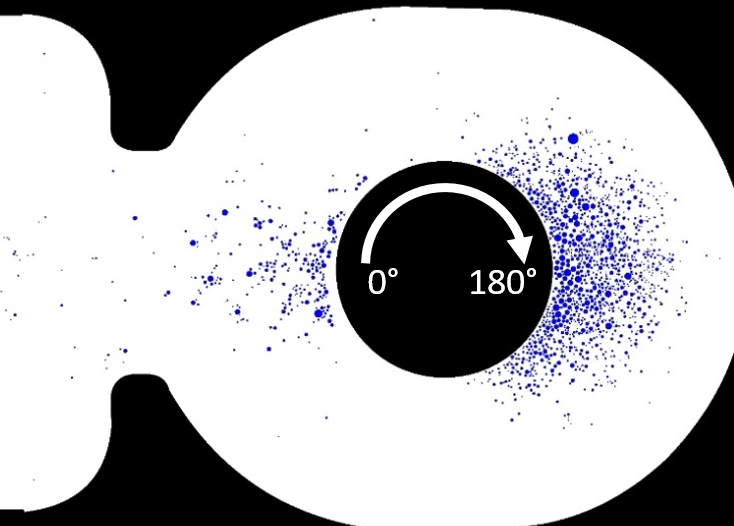}
\hspace{0.02\textwidth}
\includegraphics[width=0.5\textwidth]{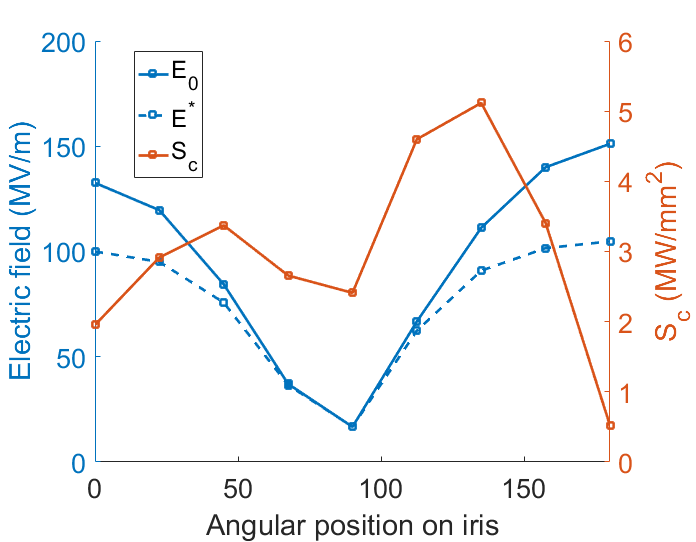}
\caption{Top: breakdown locations (represented by blue dots) vs. transverse position in  the input coupling cell of the CLIC Crab Cavity. Bottom: unloaded surface electric field (solid blue) and $E^*$ (dashed blue) in MV/m and $S_c$ (red) in MW/mm\textsuperscript{2} vs. angular position on the iris in degrees, at an input power of 43.0 MW. The definition of the angular position relative to the geometry of the cell is shown in the top plot.}
\label{fig:crab_cell1_eloaded_profile}
\end{figure}

Fig. \ref{fig:crab_cell1_eloaded_profile} shows a comparison of $E_0$, $E^*$, and $S_c$ at the locations of each of the antennas in the coupling cell, as well as a plot of actual breakdown locations in that cell. It can be clearly seen that, unlike $S_c$, $E^*$ exhibits maxima at the 0\degree and 180\degree positions, which coincides with the two clusters of breakdowns. The $E^*$ value of the 180\degree position is also higher than that in the 0\degree position, which is consistent with more breakdowns having occurred close to the former position. The magnetic field has a distribution with maxima rotated 90\degree compared to the electric field, which corresponds to the minimum of the breakdown distribution. Although the $E^*$ distribution in this cell is qualitatively similar to that of $E_0$, its absolute level is important in determining the breakdown performance of the structure, and will be discussed in more detail in Sec. \ref{sec:t24_vs_crab_results}.

\subsection{\label{sec:t24_results}CLIC T24 Structures}
A calculation was also performed on a T24 CLIC structure prototype without Higher-Order-Mode (HOM) damping \cite{T24Design}, that can be considered representative of typical undamped designs for CLIC. A large number of experimental results for this and similar geometries, but with different dimensions and RF parameters, are available, and $S_c$ has been shown to work very well for this class of structure. The coupling iris diameter of the structure is tapered, meaning among other things that the group velocity decreases from 0.018c to 0.009c along its length in order to compensate for the reduction in accelerating gradient caused by the loss of RF power to the beam and wall resistance. It has a mean accelerating gradient of 100 MV/m with no beam loading at its nominal power of 37.5 MW, and was optimised for breakdown performance using $S_c$.

\begin{figure}[htbp]
\centering
\includegraphics[width=0.5\textwidth]{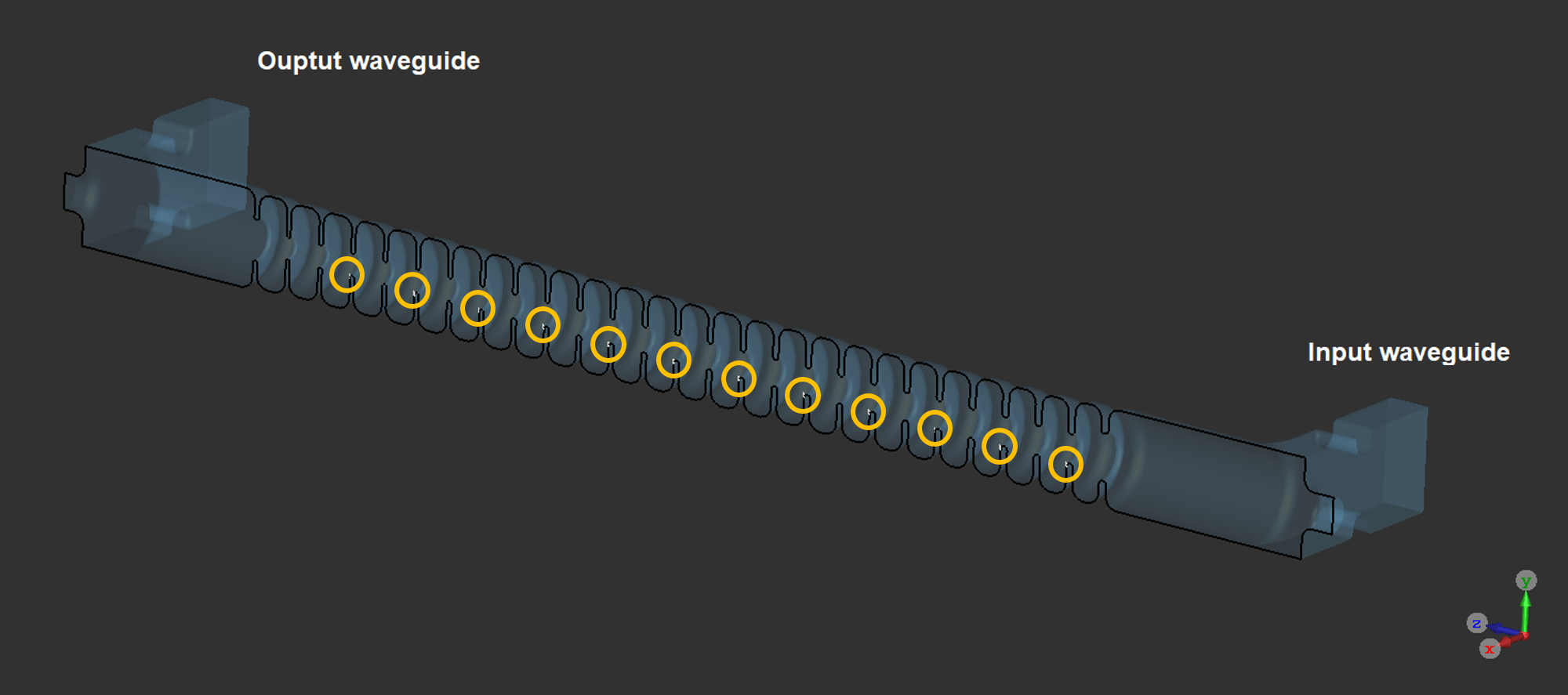}
\caption{A cutaway view of the geometry of the T24 structure with the locations of antennas shown as white lines with yellow circles around them.}
\label{fig:t24_antenna_setup}
\end{figure}

As the cells are radially symmetric, there was no need to place antennas at different azimuths. Twelve antennas were placed in various cells along the length of the structure to observe the dependence of $E^*$ on the tapering of the structure. As with the Crab Cavity, each antenna was placed slightly away from the apex of each iris at the location of maximum surface electric field.

\begin{figure}[htbp]
\centering
\includegraphics[width=0.5\textwidth]{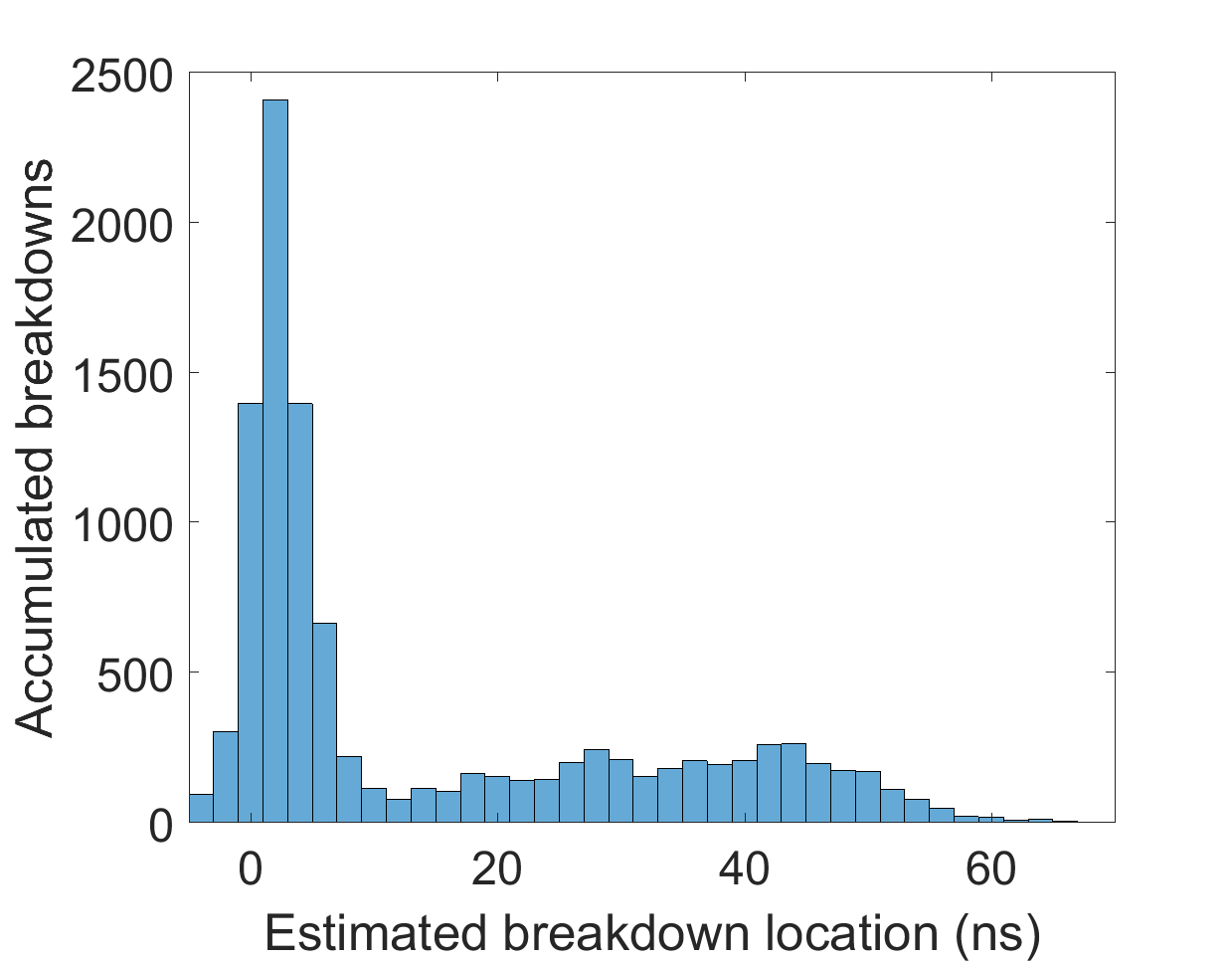}
\hspace{0.02\textwidth}
\includegraphics[width=0.5\textwidth]{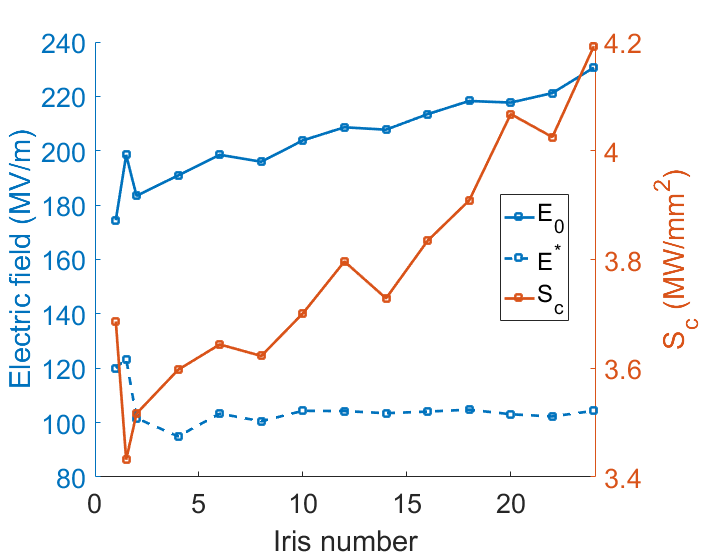}
\caption{Top: Accumulated breakdowns vs. longitudinal position in the T24PSIN1 over its entire test run in XBox 2, in units of RF signal propagation time in ns. 0 ns represents the structure input and 65 ns represents the structure output \cite{JanThesis}. Bottom: unloaded surface electric field (solid blue) and $E^*$ (dashed blue) in MV/m and $S_c$ (red) in MW/mm\textsuperscript{2} vs. iris number, at an input power of 42.4 MW.}
\label{fig:t24_eloaded_profile}
\end{figure}

The $E_0$, $E^*$, and $S_c$ values at each antenna location are compared in the bottom plot of Fig. \ref{fig:t24_eloaded_profile}. Due to the tapered design, $E_0$ and $S_c$ increase along the length of the structure and are both larger at the output end than the input end when no beam is present. Despite this, tests at the XBox test stands have shown a concentration of breakdowns near the input end of this and other structures \cite{JanThesis} \cite{TD26CC} \cite{SiC_Results}, for example as shown in the top plot of Fig. \ref{fig:t24_eloaded_profile}. On the other hand, $E^*$, whose value in the input coupler cell is significantly higher than in the regular cells, is consistent with these results.

\subsection{\label{sec:t24_vs_crab_results}Absolute $E^*$ Values}

\begin{table}[]
\caption{\label{tab:t24_vs_psi_results}Comparison of $E^*$ and $S_c$ results for the T24 structure and Crab Cavity. The reference power refers to the RF power at which each structure had a breakdown rate of $10^{-6}$ breakdowns per pulse with 200 ns pulses after conditioning.}
\centering
\begin{ruledtabular}
\begin{tabular}{l|c|c|c}
Parameter            & T24 & Crab & Units \\ \hline
Nominal power without beam & 37.5 &  13.35 & MW \\
$S_c$ at nominal power & 3.50 & 1.71 & MW/mm\textsuperscript{2} \\
$E^*$ at nominal power & 98.6 & 60.8 & MV/m \\ \hline
Reference power & 42.4 & 43.0 & MW \\
$S_c$ at reference power & 3.95 & 5.09 & MW/mm\textsuperscript{2} \\
$E^*$ at reference power & 123 & 105 & MV/m \\
\end{tabular}
\end{ruledtabular}
\end{table}

\begin{figure}[htbp]
\includegraphics[width=0.48\textwidth]{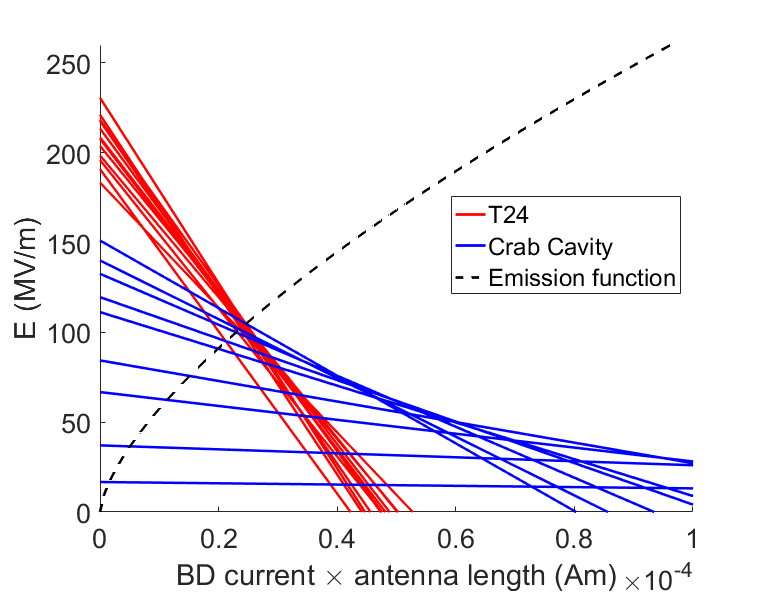}
\hspace{0.02\textwidth}
\caption{\label{fig:t24_vs_crab_fitting}Red: load lines corresponding to various regular cells of the T24 structure at 42.4 MW. Blue: Load lines corresponding to various locations inside the first cell of the Crab Cavity prototype at 43.0 MW. Black: emission function as given by \eqref{eqn:child_langmuir_emission}.}
\end{figure}

As the premise of the loaded-field model is that the $E^*$ value for a given breakdown rate and material is constant irrespective of the geometry, test results of the T24 and Crab Cavity prototype structures were used as reference values. As was the case with $S_c$, the results were compared to a pulse length of 200 ns and a breakdown rate of $10^{-6}$ breakdowns per pulse (bpp). At the end of its test run, the Crab Cavity was found to run stably under these conditions at a power level of 43 MW \cite{Woolley2015HighCavity}. The T24 structure was run with the same pulse length and a breakdown rate of $3.1\times10^{-7}$ bpp at an accelerating gradient of 103 MV/m \cite{t24_paper}. This was scaled, using the relation $E_{acc}^{30} \propto$ BDR \cite{Grudiev2009}, to $10^{-6}$ bpp and 106.3 MV/m and therefore an input power of 42.4 MW. 
If all the breakdowns in the T24 structure did indeed occur in the input matching cell after an initial conditioning period, as predicted by $E^*$ (see Sec. \ref{sec:t24_results}, the effective local breakdown rate would be much higher than the breakdown rate in the remainder of the structure, making it difficult to draw a limiting value for $E^*$ from this data. It was thus decided to consider only the regular cells of the T24 when comparing it to the Crab Cavity to pick a value for $k'$.

The load lines corresponding to the maximum $E^*$ in the Crab Cavity and the regular cells of the T24 cross the point ($E^* =$ 105 MV/m, $I^*l_{ant} = 2.47\times 10^{-5}$ Am), as shown in Fig. \ref{fig:t24_vs_crab_fitting}. The value of $k'$ was chosen such that the emission function crossed this point, thus ensuring agreement in their maximum $E^*$ values. It was decided that this should be considered equivalent to the mean of the $S_c$ values of the two structures at their reference power levels, i.e. $S_c = $ 4.52 MW/mm\textsuperscript{2}.

Results related to the peak $E^*$ and $S_c$ values in the two structures can be found in Table \ref{tab:t24_vs_psi_results}. Although there were significant differences in the design of the two structures, they were made of the same material and a breakdown quantity should have the same value in both structures when fully conditioned and operating at the same breakdown rate and pulse length. Comparing reference values, $S_c$ differed by 28.7\% between the two structures, implying a 13.4\% discrepancy in predicted maximum field values. 

If the input matching cell of the T24 is disregarded, the $E^*$ values of the two structures at their reference power levels are identical by virtue of the choice of $k'$. If the matching cell is taken into account, the maximum $E^*$ value in the T24 rises to 123 MV/m. This is a 17.5\% discrepancy in $E^*$ and a 21.4\% discrepancy in predicted unloaded field. It should also be noted that the $k'$ and threshold $E^*$, and the corresponding $S_c$ value used here also exhibited good agreement for the C3 structures as shown in Fig. \ref{fig:c3_max_gradient_transient}.

\section{\label{sec:extensions}Additional Considerations}
\subsection{\label{sec:broadband_vs_single_freq}Comparison of Single-Frequency and Broadband Results}

\begin{figure}[htbp]
\centering
\includegraphics[width=0.49\textwidth]{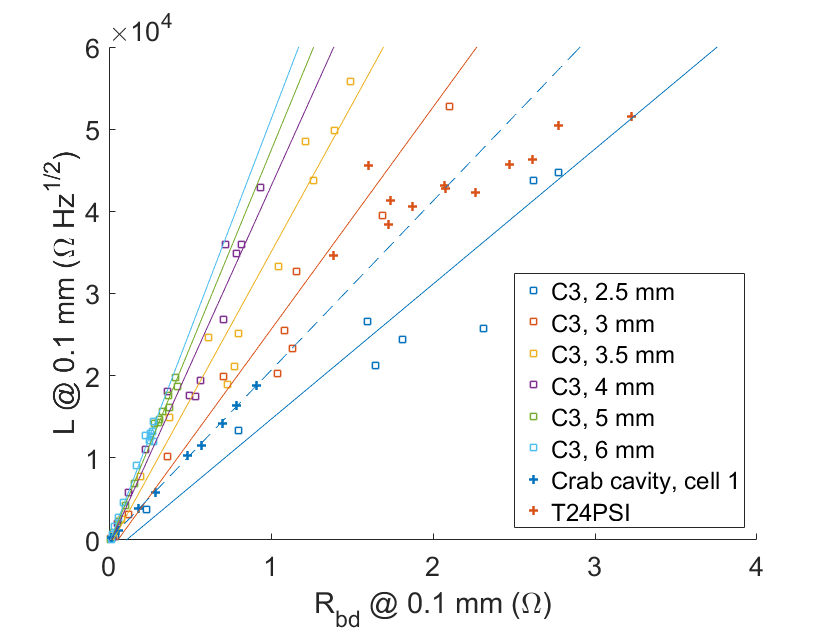}
\caption{$L$ vs. single-frequency $R_{bd}$ for C3 structures of different aperture sizes, as indicated on the legend, scaled to an antenna length of  0.1 mm. Results for the T24PSI and the first cell of the CLIC Crab Cavity, scaled to an antenna length 0.1 mm, are also shown. Solid lines represent linear fits to the C3 structure data, whereas the blue dashed line represents a linear fit to the data from the Crab Cavity.}
\label{fig:all_structures_rbd_transient_vs_rbd}
\end{figure}

\begin{figure}[htbp]
\centering
\includegraphics[width=0.48\textwidth]{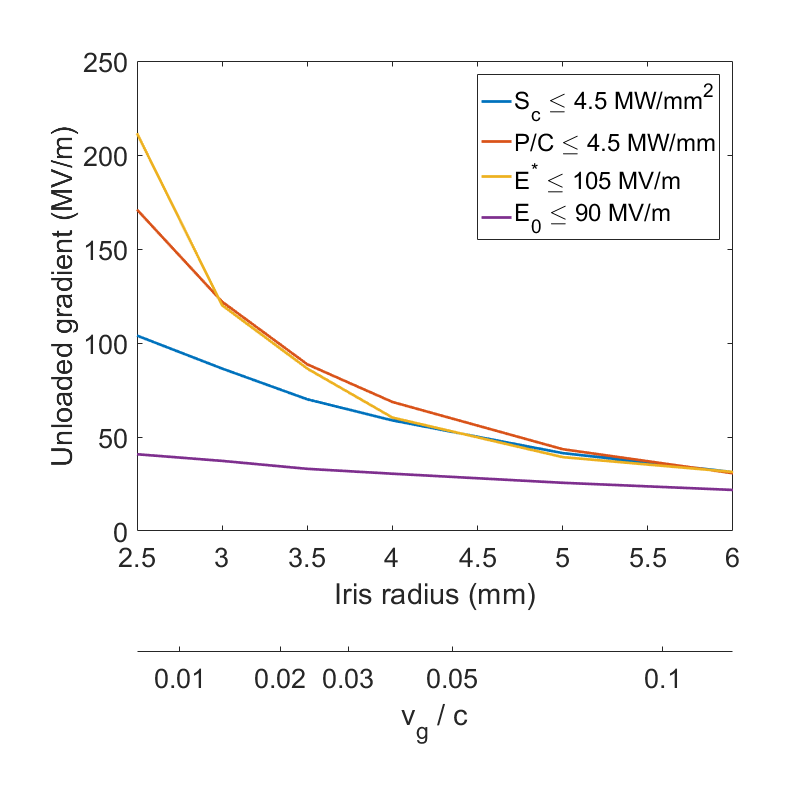}
\caption{Maximum accelerating gradient without beam loading vs. aperture size and group velocity as a fraction of $c$. Blue curve: with a maximum permitted $S_c$ of 4.5 MW/mm\textsuperscript{2}. Red curve: with a maximum permitted $P/C$ of 4.5 MW/mm. Yellow curve: with a maximum permitted $E^*$ of 105 MV/m.}
\label{fig:c3_max_gradient}
\end{figure}

To investigate the difference between the broadband and single-frequency methods, the values of $R_{bd}$($f = 12$ GHz) and $L$ for all of the antenna locations in all of the numerical simulations discussed in this paper were compared and are shown in Fig.~\ref{fig:all_structures_rbd_transient_vs_rbd}. For each of the aperture sizes of the C3 structure, as well as the Crab Cavity, an approximate dependence $L \propto R_{bd}$ can be seen, whose slope depends of the bandwidth of the structure, which is related to the group velocity for travelling-wave structures. The C3 structures with larger aperture sizes and thus larger bandwidths correspond to a steeper slope on the plot. Each of the data points for the T24 structure have different ratios of $L$ to $R_{bd}$, reflecting the tapered design of the structure.

Fig. \ref{fig:c3_max_gradient} shows the predicted maximum accelerating gradient for C3 structures of various aperture sizes, with $E^*$ calculated using the $R_{bd}$ value at 12 GHz instead of $L$, with a constant $k'$ value. The $E^*$ curve diverges from the $S_c$ curve significantly for low-group-velocity structures, demonstrating the issues with using the single-frequency method with narrowband structures. The agreement between the curves is much better for larger group velocities, indicating that the single-frequency method might be sufficient for such structures.

\subsection{\label{sec:freq_dependence}Scale Dependence}
A good limiting quantity should give consistent results for a wide range of frequencies without the need for re-fitting the free parameters separately for each frequency. $S_c$ has been found to be in general agreement with experimental results for 12 GHz and 30 GHz accelerating structures \cite{Grudiev2009}.

To investigate the dependence on frequency, the effect of scaling a given geometry by a factor $s$ and the RF frequency $f$ correspondingly by a factor $1/s$ will be considered. The field patterns in this geometry, normalised to accelerating gradient and scaling, i.e. $\mathbf{E}(\vec{x}/s)/E_{acc}$ and $\mathbf{H}(\vec{x}/s)/E_{acc}$, where $\vec{x}$ is the position inside the geometry, remain invariant as a consequence of the scale invariance of Maxwell's equations \cite{MicrowaveEngineering}. When normalised in a similar fashion, $S_c(\vec{x}/s) / E_{acc}^2$ is also scale-invariant. This means that the maximum gradient for a given maximum $S_c$ value is scale-invariant.

The same logic can be applied to fP/C: the normalised Poynting vector $\mathbf{S}(\vec{x}/s) / E_{acc}^2$ is scale-invariant. Thus, the total power $P$ flowing into the structure, given by integral of the real part of the axial vector component of the Poynting vector over the area of the iris, i.e. $\int_{\text{iris}}\Re(S_z)dA$, scales as $s^2$. Since the iris circumference $C$ scales as $s$, $fP/C$ is scale-invariant.

The dependence of $E^*$ on scaling can be determined from the equation for $E^*$:

\begin{equation}
\label{eqn:eloaded_solution_freqscaling}
    k l_{ant}^{1/2} L_0 E^{*3/2} + E^* - E_0 = 0.
\end{equation}

Since $k$ should only describe properties of the emission site, it should not depend on $s$. As discussed in Sec.~\ref{sec:antenna_lengths}, it is reasonable to assume $l_{ant}$ should be scaled proportionally to the geometry by factor $s$. As can be deduced from Eqs. \eqref{eqn:analytical_rbd} and \eqref{eqn:square_integral_rbd}, $L_0$ is proportional to the square root of bandwidth, which is in turn proportional to the operating frequency, i.e. $L_0 \propto s^{-1/2}$. This dependence cancels out that of $l_{ant} \propto s$, resulting in the coefficients of $E^*$ in Eq.~\eqref{eqn:eloaded_solution_freqscaling}, and therefore the $E^*$ value that satisfies the equation, being scale-invariant.

The fact that all three quantities are independent of scale suggests that $E^*$ should be valid for RF cavities of other frequencies than 12 GHz without modification, given that $E^*$ and $S_c$ have been shown to show similar results for various $TM_{010}$ geometries.

\subsection{\label{sec:exponent_dependence}Dependence on the Exponent of the Emission Function}

\begin{figure}[htbp]
\centering
\includegraphics[width=0.235\textwidth]{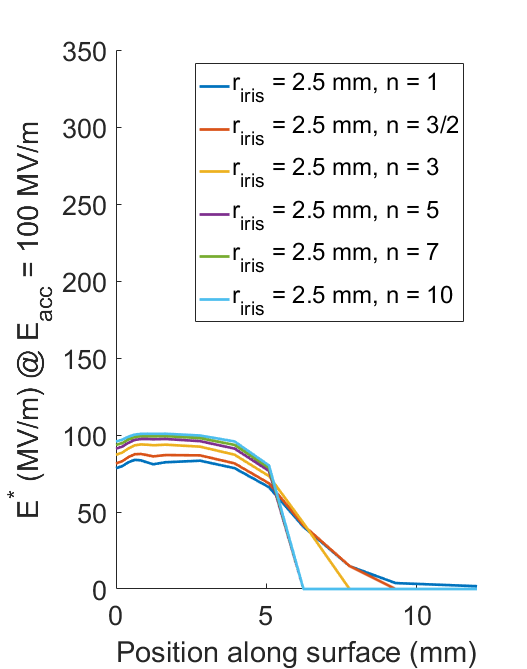}
\includegraphics[width=0.235\textwidth]{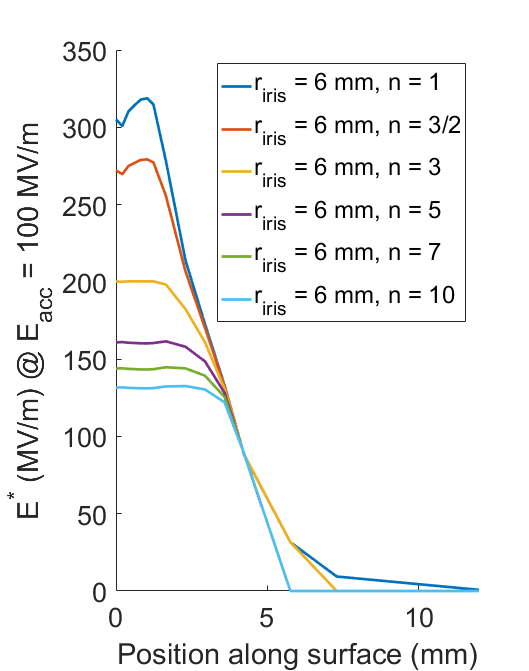}
\caption{$E^*$ at an accelerating gradient of 100 MV/m vs. position along the surface in mm. The position is defined by the arrow shown in Fig.~\ref{fig:c3_dimensions}, where 0 is the iris apex, for different values of exponent $n$ of the electric field term in the emission function. Left: For an aperture radius of 2.5 mm. Right: For an aperture radius of 6 mm.}
\label{fig:n_sweep}
\end{figure}

Although a justification for the use of the Child-Langmuir law as the emission function was given in Sec. \ref{sec:emission_function}, it may not describe exactly the behaviour of real breakdowns. Reasons for this include the fact that the emitted electrons are not constrained to move in one dimension, as well as relativistic effects. The actual emission function may have an electric field term with a different exponent from the 3/2 as assumed in Eq. \eqref{eqn:child_langmuir_emission}, or have a different functional form altogether.

To verify that the $E^*$ method has some level of robustness to uncertainty in the emission function, calculations of $E^*$ were done for the C3 structures with various exponent values $n$ between 1 and 10 in the emission function. In each case a value of $k'$ was chosen such that the function crossed the point ($E^* =$ 105 MV/m, $I^*l_{ant} = 2.47\times 10^{-5}$ Am). No differences in the maximum predicted gradient vs. aperture radius, as shown in Fig. \ref{fig:c3_max_gradient_transient}, were found for the exponent values tried. Differences in the spatial distribution of $E^*$ on the surface of the cell were noted, however, as shown in Fig. \ref{fig:n_sweep}. For larger values of $n$, the spatial distribution of $E^*$ became flatter across the high-electric-field region, with a sharper drop off to zero outside it, since the larger the exponent value, the closer the emission function becomes to a sharp threshold in $E$. For exponent values of 5 and above, the maximum values of $E^*$ for the 6 mm aperture radius case no longer coincided with the maximum $E_0$, due to the effect of lower $L$ in those regions outweighing the smaller $E_0$ value.

The $E^*$ results for the C3 structures thus appear good for exponent values between 1 and 3, showing a degree of robustness to differences in the emission function.

\subsection{\label{sec:thin_gaps}Behaviour in Thin Parallel-Plate Geometries}

Setups featuring a thin gap between two parallel conducting surfaces are common in the study of vacuum breakdown, particularly in the case of DC experiments, as well as choke-mode RF structures \cite{Shintake_Choke}. A dependence of the maximum electric field on the gap size has been noted in experiments involving both RF \cite{CLIC_Choke_Design} and DC electric fields \cite{gap_dependence_histogram}. 

To investigate the behaviour of breakdown impedance in such a setup, one can consider the case of an antenna spanning such a gap. As discussed in Sec. \ref{sec:antenna_lengths}, the electron trajectory, and therefore the antenna length, is assumed to be constrained by the thickness of the gap, i.e. $l_{ant} = d$, with $d$ being the gap thickness. Assuming that the thickness of the gap is much smaller than the free-space RF wavelength $\lambda_{fs}$, one can assume that there is no variation in the electric or magnetic fields with distance normal to the surfaces. This means that the field patterns are symmetric about the plane halfway between the two electrodes, and a conductive plane could therefore be inserted in place of this symmetry plane without perturbing the field patterns. This means that the perturbation in electric field $\Delta E$ caused by a current $\Delta I$ in the antenna is the same for a gap size of $d$ or $d/2$ for any $d \ll \lambda_{fs}$. One can therefore deduce that $\Delta E/\Delta I$ is independent of the gap size, and that $R_{bd} \propto d$. The quadratic scaling law of Eq. \eqref{eqn:r0_definition} does not apply here as the geometry changes with $d$ and thus $R_0$ cannot be assumed constant. This implies:

\begin{equation}
    R_0 = \frac{R_{0,n}}{d},
\end{equation}

with $R_{0,n}$ being the $R_0$ value for unit $d$. Substituting this into Eq. \eqref{eqn:eloaded_solution}:

\begin{equation}
\label{eqn:eloaded_solution_gap_scaling}
    E_0 = k R_{0,n} E^{*3/2} d^{-1/2} + E^*.
\end{equation}

It should be noted that with very small gap sizes of 1 mm of less, the emission function is very likely to be the Child-Langmuir law due to 1D electron trajectories being much more likely, and voltage being too low (assuming electric field levels on the order of 100 MV/m) to cause significant relativistic effects.

Eq. \eqref{eqn:eloaded_solution_gap_scaling} implies that for very large gaps, $\lim_{d \to \infty} E_0 = E^*$, whereas for small gaps, $E_0$ can be much higher than $E^*$. This supports the intuition that a very small gap can restrict the flow of power to a breakdown and allow higher operating electric fields. This equation can be related to a study of the gap size dependence performed in the DC Large Electrode System (LES) at CERN \cite{Iaroslava_MeVArc_2019}. In this experiment, it was found that for a constant breakdown rate, the gap voltage varied as $V_0 \propto d^{0.72}$. This can be compared with the loaded-field model by expressing Eq. \eqref{eqn:eloaded_solution_gap_scaling} in terms of gap voltage as:

\begin{equation}
\label{eqn:eloaded_solution_gap_scaling_voltage}
    V_0 = E_0 d = k R_{0.n} d^{1/2} + E^* d.
\end{equation}

The constant breakdown rate in the experiment implies a constant value of $E^*$. It is likely that with appropriate values of the coefficients $k$, $R_{0,n}$, and $E^*$, a linear combination of the $d$ and $d^{1/2}$ terms would be indistinguishable from a power law with exponent 0.72.

Other experiments involving parallel-plate and near-parallel-plate geometries \cite{gap_dependence_histogram} have shown exponents between 0.1 and 1.1, with 0.7 being the most common. Eq. \eqref{eqn:eloaded_solution_gap_scaling_voltage} can explain a spread of exponent values between 0.5 and 1 by differences in the relative values of $k R_{0,n} E^{*3/2}$ and $E^*$ in each setup.

\section{\label{sec:implications}Implications on Accelerating Structure Design}
The $E^*$ model has implications on the behaviour of vacuum breakdowns and the design of structures for optimum breakdown performance. A key feature of the $E^*$ model is that the breakdown rate depends only on the local electric field, and not the magnetic field, though there is generally a correlation between power coupling and magnetic field levels. $E_0$ should thus be minimised where possible to avoid breakdowns.

A $k'$ value of $2.30\times10^{-17}$ Am\textsuperscript{2}V\textsuperscript{-3/2} and an $E^*$ of 105 MV/m have been shown to produce results consistent with an $S_c$ value of 4.52 MW/mm\textsuperscript{2}. Lower vales of $S_c$ correspond to the same $k'$ with a lower $E^*$. The broadband method is recommended in general, though the single-frequency model could suffice for designs with a sufficiently large group velocity and not much variation in group velocity, offering the advantage of simpler calculations.

Exotic designs such as structures driven with multiple frequencies to reduce the peak surface field have been proposed \cite{MultiFrequencyStructure}. Based on the premise of the $E^*$ model, this could be effective in improving performance, though the model may need modification to handle multi-frequency excitation.

The $E^*$ model shows an explicit dependence of accelerating gradient on group velocity - a correlation which has been observed experimentally and noted in previous work \cite{Adlophsen_Normal_Conducting} \cite{SLAC_Gradient_Limit}.

The behaviour of the T24 structure in Sec.~\ref{sec:t24_results}, i.e.: $E^*$ and the breakdown rate being the largest in the cell closest to the input and thus receiving the most RF power, suggests that coupling power to each cell individually, such as in the distributed-coupling structure designed at SLAC \cite{TantawiDistributedCoupling}, may improve breakdown performance. This way, a breakdown in any particular cell would only have a small fraction of the total input RF power to the whole structure available to it, which could bring about an improvement in breakdown performance. Local stored energy would still needs to be taken into account in such a design.

\section{\label{sec:conclusions}Conclusions}
A new model to quantify breakdown limits in normal-conducting accelerating structures has been proposed. It describes the coupling of power to a breakdown, and takes into account the perturbation in local $E$ and $H$ fields by the breakdown whilst also removing the intricacies of calculating actual particle trajectories for every possible breakdown location.

$E^*$, the quantity that predicts the probability of a breakdown of a particular geometry at a given input power, is relatively simple to calculate and can be obtained either analytically from standard cavity parameters or numerically from finite-element simulations.

The model resolves issues with the breakdown locations predicted by $S_c$ whilst still predicting maximum gradient values that are consistent with it for geometries that $S_c$ has been well-tested for (i.e. accelerating structures operating in the $TM_{010}$ mode). Like $S_c$, the loaded-field model takes stored energy into account and is thus valid for structures with low group velocities. It also exhibits scale invariance similarly to $fP/C$ and $S_c$, meaning that it should be equally applicable to different frequency cavities without changes of parameters. Because the $E^*$ model does not require steady-state power flow, it is in principle also applicable to DC systems. Although standing-wave structures with zero group velocity have not been considered directly in this paper, obtaining an analogous analytical circuit model to those discussed in Sec. \ref{sec:analytical_rbd} for such structures is expected to be straightforward, while the numerical approach discussed in Sec. \ref{sec:numerical_rbd} can be applied without modification.


\bibliography{references}
\end{document}